\documentclass[prl,twocolumn,superscriptaddress,amsmath,amssymb,floatfix,noshowpacs]{revtex4}
\usepackage{graphicx,hyperref}
\usepackage{color}
\usepackage{soul}
\graphicspath{{fig/}}

\newcommand*{\vect}[1]{\mathbf{#1}}
\newcommand*{\bra}[1]{\langle #1 \vert}
\newcommand*{\ket}[1]{\vert #1 \rangle}
\renewcommand*{\tensor}[1]{\overline{\overline{{#1}}}}
\newcommand*{\comm}[2]{\left[#1, #2\right]}

\newcommand{\hide}[1]{}

\newcommand{\eq}[1]{Eq.\,(\ref{#1})}

\newcommand{\noeq}[1]{(\ref{#1})}
\newcommand{\fig}[1]{Fig.\,\ref{#1}}

\newcommand{\nofig}[1]{\ref{#1}}

\begin{document}
\title{\textbf{Cooperative resonances in light scattering from two-dimensional atomic arrays}}
\author{Ephraim Shahmoon}
\affiliation{Department of Physics, Harvard University, Cambridge MA 02138, USA}
\author{Dominik S.~Wild}
\affiliation{Department of Physics, Harvard University, Cambridge MA 02138, USA}
\author{Mikhail D.~Lukin}
\affiliation{Department of Physics, Harvard University, Cambridge MA 02138, USA}
\author{Susanne F.~Yelin}
\affiliation{Department of Physics, Harvard University, Cambridge MA 02138, USA}
\affiliation{Department of Physics, University of Connecticut, Storrs, Connecticut 06269, USA}
\date{\today}

\begin{abstract}
  We consider light scattering off a two-dimensional (2D) dipolar array and show how it can be tailored by properly choosing the lattice constant of the order of the incident wavelength. In particular, we demonstrate that such arrays can operate as nearly perfect mirrors for a wide range of incident angles and frequencies close to the individual atomic resonance. These results can be understood in terms of the cooperative resonances of the surface modes supported by the 2D array.
  Experimental realizations are discussed, using ultracold arrays of trapped atoms and excitons in 2D semiconductor materials, as well as potential applications ranging from atomically thin metasurfaces to single photon nonlinear optics and nanomechanics.
\end{abstract}

 \maketitle
Control over propagation and scattering of light fields plays a central role in optical science and photonics. In particular, it is well known that emitters exhibit a strongly modified linear and nonlinear optical response on resonance. For example, enhanced optical scattering in 2D arrays of linearly polarizable elements have been extensively studied in photonics \cite{EBS,GDA1,MOR,GDA2}. Recently, it has been shown  that thin 2D metamaterials, known as metasurfaces, whose constituent elements are optical antennas with varying resonances, can drastically alter the transmitted field by enabling spatial control of its amplitude, phase and polarization~\cite{CAP,SHA}.
As a rule, these elements are micro-fabricated from macroscopic material, while the separation between the array elements is typically much smaller than the operating wavelength. At the same time,  resonant light can be completely reflected by individual atoms when they are strongly coupled to nanophotonic devices with sub-wavelength localization of light \cite{CHA,FAN,ASE,LAU,APP}. Intuitively, this originates from resonant enhancement of the optical cross section of a polarizable dipole, which at resonance universally scales as $\lambda^2$, $\lambda$ being its resonant wavelength. Such single atom reflectors yield extraordinary nonlinearities at the level of individual photons \cite{LUKT,DAY,ALP}.

\begin{figure}[t]
  \begin{center}
    \includegraphics[width=\columnwidth]{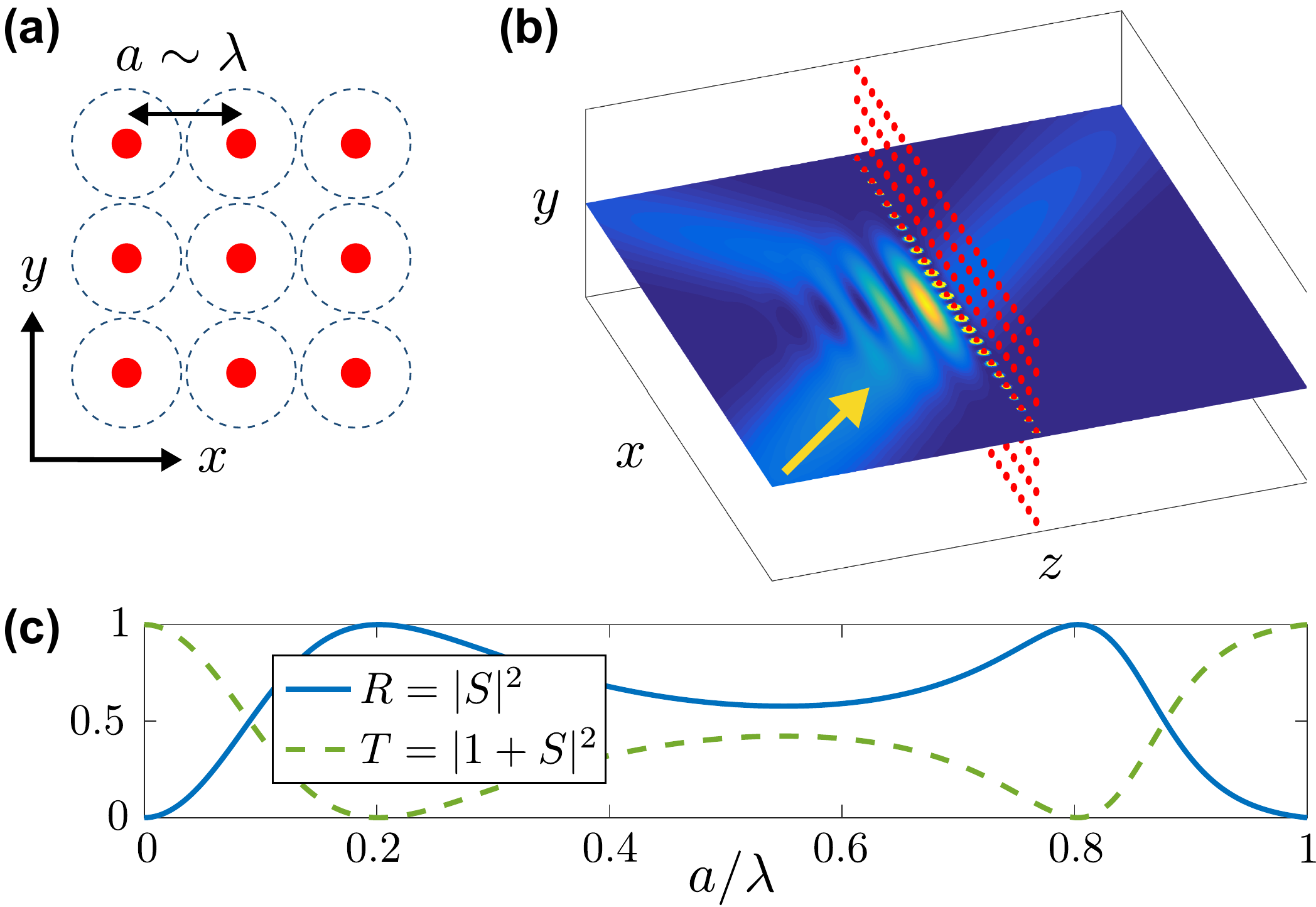}
    \caption{\small{
      (a) 2D array of atoms spanning the $xy$ plane at $z=0$, with inter-atomic spacing $a$ on the order of the resonant wavelength of the atoms, $\lambda$. For resonant light, the individual atomic cross section is of order $\lambda^2$ (dashed circles).
      (b) Light scattering off the array in the single diffraction order regime: The incident field (yellow arrow) produces a forward scattered field at $z>0$ and a reflected field at $z<0$.
      (c) Intensity transmission coefficient $T$ and reflection coefficient $R$ for a square-lattice at normal incident and resonant light ($\delta=0$) as a function of the lattice constant $a$. Strong scattering is observed with perfect reflection (vanishing transmission) occurring at $a/\lambda \approx 0.2,0.8$.
    }} \label{fig1}
  \end{center}
\end{figure}

In this Letter we explore light scattering from a 2D ordered and dilute array of atoms, with a lattice constant of the order of a wavelength, as can be realized, e.g. using ultracold atoms
loaded into optical lattices~\cite{OL,BRO}. In such a case  near-resonant operation can still lead to strong scattering. Indeed, vanishing transmission at normal incidence was recently discovered in a numerical study of a 2D atomic lattices for a specific frequency and lattice arrangement \cite{ADM1}. Due to resonant enhancement, one may na\"{\i}vely expect that a single layer of dipoles, even if they are as small as individual atoms, may ``tile'' the plane and thus act as a strong scatterer, provided the density of dipoles exceeds $1/\lambda^2$ (\fig{fig1}a). This reasoning, though providing intuition for the possibility of strong scattering in dilute media, ignores the important effect of interactions between the dipoles \cite{RUO,KUP,SCU,ROB}.

In what follows, we develop an analytical approach to the scattering problem and show that the near-unity reflection is a generic phenomenon associated with  the \emph{cooperative resonances} of the dipolar array and its
collective surface-wave excitations. Strong scattering  occurs when the frequency of the incident light matches that of the cooperative resonance of the array.
The control of scattering off the array can be achieved by adjusting the lattice constant, which determines the cooperative resonances. We demonstrate that the array can form a perfect mirror at almost all incident angles and polarizations, where both the resonant frequency and the bandwidth can be tuned.

\emph{The atomic array.---}
We consider a 2D array of identical point-like particles with a generic linear and isotropic polarizability \cite{PET}
\begin{equation}
  \alpha(\delta)=-\frac{3}{4\pi^2}\epsilon_0\lambda_a^3\frac{\gamma/2}{\delta+i(\gamma+\gamma_\mathrm{nr})/2}.
  \label{alp}
\end{equation}
Here $\delta=\omega-\omega_a$, with $|\delta|\ll\omega_a$, is the detuning between the frequency of the incident light $\omega=2\pi c/\lambda$ and that of the resonance of the particles $\omega_a=2\pi c/\lambda_a$, and $\gamma$ ($\gamma_\mathrm{nr}$) is the radiative (non-radiative) width of this resonance. For a closed cycling transition in atoms we have $\gamma_\mathrm{nr}=0$ and the isotropic and linear response corresponds to a $J=0$ to $J=1$ transition far from saturation. The array is taken to be an infinite square lattice with lattice constant $a<\lambda$, spanning the $xy$-plane at $z=0$ (\fig{fig1}a).

\emph{Scattering at normal incidence.---}
We first focus on the simplest case of a plane wave at normal incidence. The condition $a<\lambda$ guarantees that only a single diffraction order is present in the far field such that the scattered field on both sides of the array consists of plane waves propagating in the $z$-direction (\fig{fig1}b for waves along $z$). Figure \nofig{fig1}c shows the transmission and reflection coefficients as a function of the lattice constant, computed for resonant light $\delta = 0$ and in the absence of non-radiative losses, $\gamma_\mathrm{nr} = 0$, using our analytical approach presented below. We observe that the array scatters strongly over a wide range of lattice constants. In particular, complete reflection (zero transmission) is observed at lattice constants $a/\lambda \approx 0.2$, $0.8$. We note that the null transmission at $a/\lambda \approx 0.8$ was also recently found numerically in Ref.~\cite{ADM1}.

Let us now analyze the above situation. For $a<\lambda$ the total field can be written as
\begin{equation}
  \mathbf{E}=\left[e^{ikz}+Se^{ik|z|}\right]\mathbf{E}_0,
  \label{En}
\end{equation}
where $\mathbf{E}_0$ is the amplitude of the field polarized in the $xy$-plane, $k=\omega/c$, and $S$ is a scattering amplitude. For $S=-1$, the transmitted field (at $z>0$) vanishes and the corresponding perfect reflection gives rise to a standing wave for $z<0$. The scattering amplitude is determined by the polarization $\mathbf{p}$ induced on the atoms by the incident field, which is identical for all atoms in this case. In turn, $\mathbf{p}$ is the result of multiple scattering of the incident field by all atoms in the array, and it can be characterized by an \emph{effective polarizability} of the atoms defined by $\mathbf{p}=\alpha_{e}(\delta)\mathbf{E}_0$. A self-consistent solution of this multiple-scattering problem yields \cite{SM}
\begin{equation}
  S(\delta)=i \pi\left(\frac{\lambda}{a}\right)^2\frac{\alpha_{e}(\delta)}{\varepsilon_0\lambda^3}
  =-\frac{i(\gamma+\Gamma)/2}{\delta-\Delta+i(\gamma+\gamma_{\mathrm{nr}}+\Gamma)/2}.
  \label{Sn}
\end{equation}
By comparing the structure of this linear response to that of an individual atom, \eq{alp}, we infer that the dipolar interaction between atoms in the array renormalize both the width $\gamma$ and the resonant frequency $\omega_a$. They are now supplemented by their cooperative counterparts $\Gamma$ and $\Delta$, respectively, given by
\begin{equation}
  \Delta-\frac{i}{2}\Gamma=- \frac{3}{2}\gamma \lambda \sum_{n\neq 0} G(0,\mathbf{r}_n), \quad \Gamma=\gamma\frac{3}{4\pi}\left(\frac{\lambda}{a}\right)^2-\gamma.
  \label{Gn}
\end{equation}
Here, $G(0,\mathbf{r}_n)$ is the transverse component ($xx$ or $yy$) of the dyadic Green's function of electrodynamics in free space \cite{NH}, evaluated between the central atom (``$n=0$'') at $\mathbf{r}_0=0$ and the atom $n$ at $\mathbf{r}_n$. The explicit expression for $\Gamma$ holds for $a<\lambda$ and is in fact valid for \emph{any} 2D lattice \cite{SM}.

\begin{figure}[t]
  \begin{center}
    \includegraphics[width=\columnwidth]{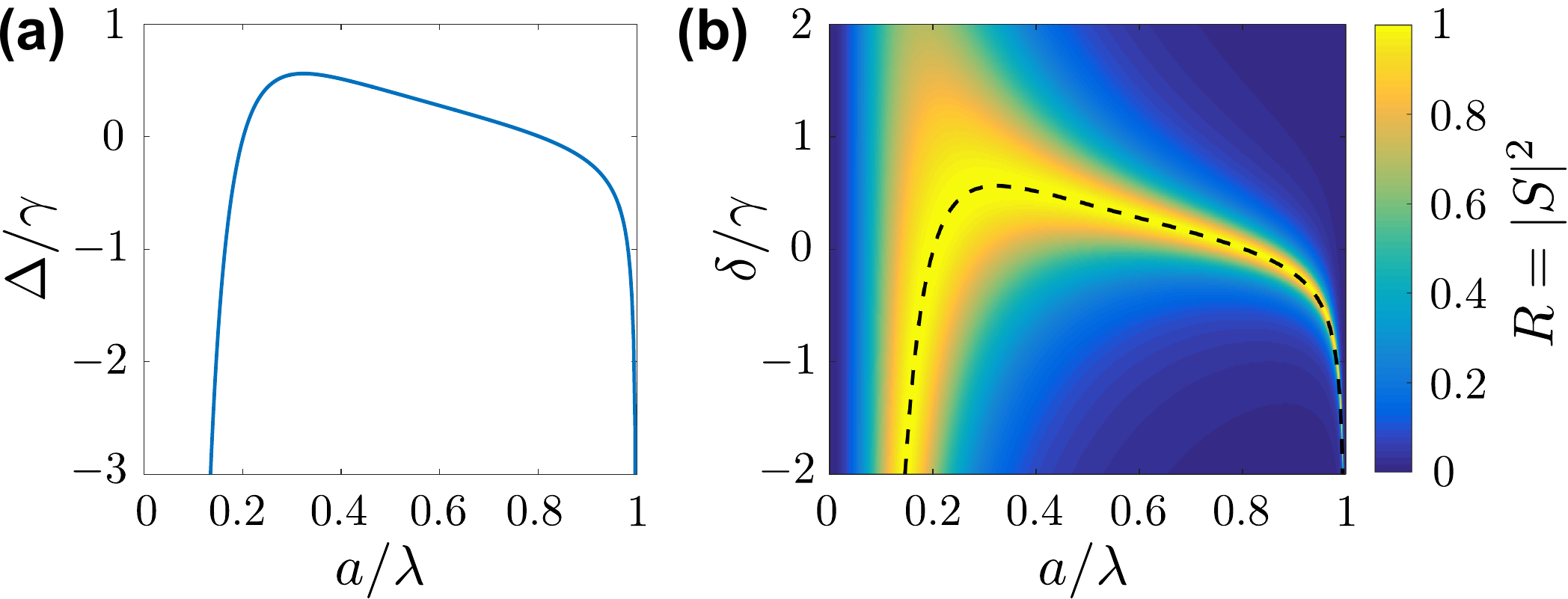}
    \caption{\small{
      (a) The cooperative shift $\Delta$, \eq{Gn}, as a function of the lattice constant $a$ (normal incidence). This plot is central in the design of the scattering since the shift determines the collective resonances of the array according to \eq{Sn}. Perfect reflection occurs when the cooperative shift equals the incident detuning, $\delta=\Delta$. For example, $\Delta=0$ at $a/\lambda \approx 0.2$, $0.8$  explains the resonances in \fig{fig1}c. (b) Intensity reflection coefficient $R$ as a function of lattice constant $a$ and detuning $\delta$. We note that the emerging contour of perfect reflection (bright yellow) coincides with the cooperative resonance plotted in (a) (marked here by the dashed black curve).
    }}
    \label{fig2}
  \end{center}
\end{figure}

Equation \noeq{Sn} reveals that scattering is strongest when the frequency of the incident light matches the cooperative resonance, $\delta = \Delta$. Perfect reflection ($S=-1$) occurs  if additionally $\gamma_\text{nr} = 0$.
Therefore, the \emph{key ingredient} that determines the scattering properties of the array is the cooperative dipole-dipole shift $\Delta$, given by the summation (readily evaluated numerically) of the dispersive dipole-dipole shift over all atoms, the real part of \eq{Gn}. Figure \nofig{fig2}a provides us with a central tool by which to understand and design the scattering off the array, as it presents the cooperative shift $\Delta$ as a function of the lattice constant $a$ \cite{SM}. For example, the vanishing cooperative shift $\Delta$ at $a/\lambda\approx 0.2,0.8$ explains the perfect reflection obtained in \fig{fig1}c for  $\delta=0$.
Moreover, \fig{fig2}a shows that scattering resonances exist for a wide range of incident field detunings $\delta$ near the individual-atom resonance. This is illustrated by \fig{fig2}b, in which the reflection coefficient is plotted as a function of both $a$ and $\delta$.

For lossy particles, where $\gamma_\mathrm{nr}\neq 0$, the scattering amplitude (\ref{Sn}) at resonance becomes $S=-(\Gamma+\gamma)/(\Gamma+\gamma+\gamma_\mathrm{nr})$. Therefore, high reflection requires that radiation damping via scattering is dominant over all other damping sources, $\gamma+\Gamma\gg\gamma_\mathrm{nr}$. The scaling $\gamma + \Gamma \propto (\lambda/a)^2$, originating from cooperative enhancement, then implies that this can be achieved for a sufficiently small lattice constant even if the individual dipoles are poor radiators ($\gamma < \gamma_\mathrm{nr}$).

\begin{figure}[t]
  \begin{center}
    \includegraphics[width=\columnwidth]{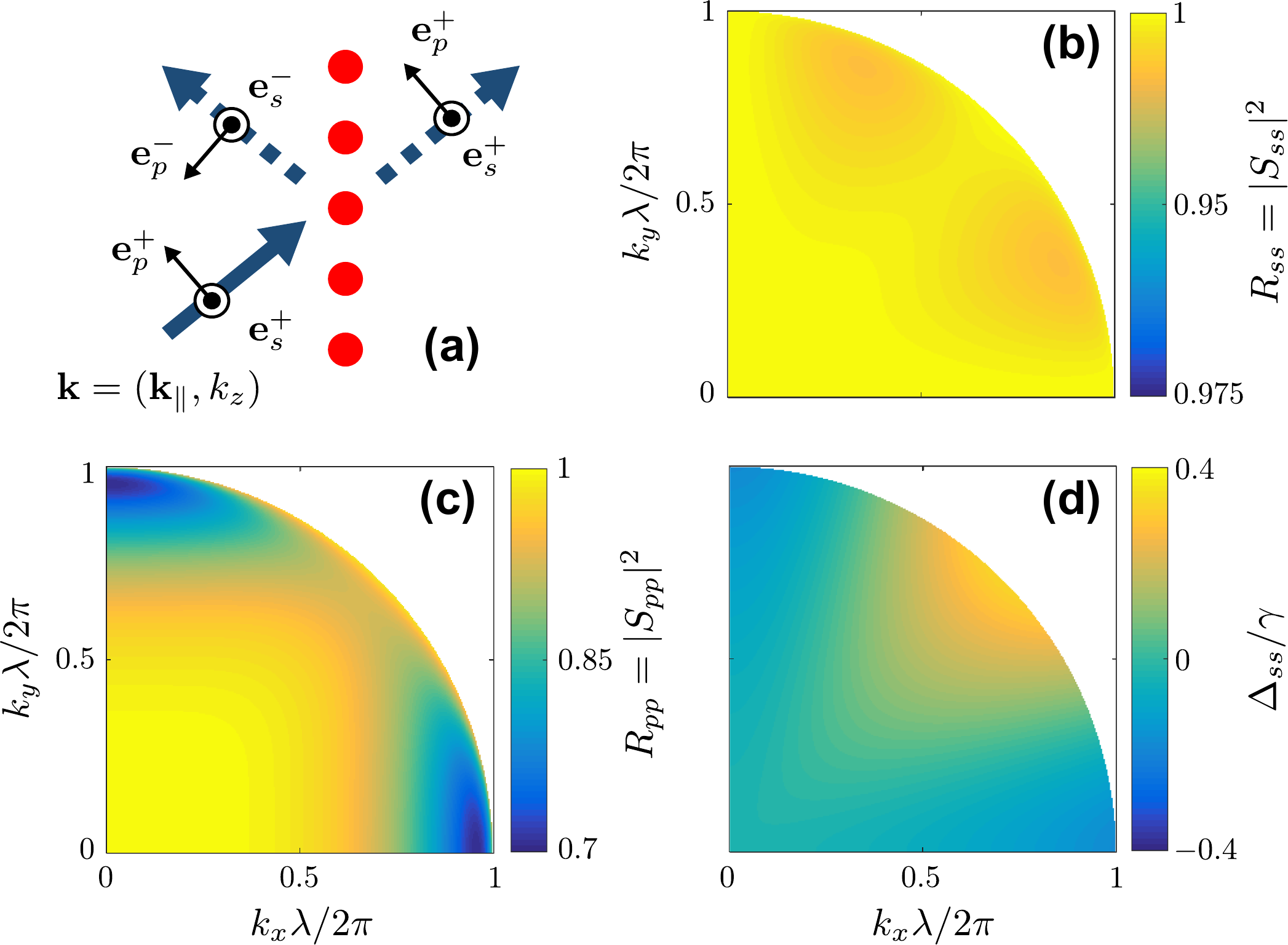}
    \caption{\small{
      Scattering at a general angle of incidence. (a) The polarizations of incident and forward scattered fields with wavevector $\mathbf{k}$ are spanned by the forward basis $\mathbf{e}^+_{p,s}\bot \mathbf{k}$, whereas that of the reflected field is spanned by the backward polarization basis $\mathbf{e}^-_{p,s}\bot\mathbf{k}$ \cite{SM}.
      (b) Scattering properties of an array with lattice constant $a=0.2\lambda$. Intensity reflection coefficient $R_{ss}$ for $s$-polarized incident and scattered fields at zero detuning from the bare atomic resonance ($\delta=0$) as a function of the in plane components $k_{x,y}$ of the incident wavevector. (c) Same as (b) for $p$-polarized reflected and incident fields. (d) $ss$ component of the cooperative shift matrix $\tensor{\Delta}$. The variation of the energy shift around the resonance $\Delta_{ss}=\delta=0$ is small compared to an atomic linewidth, which explains the high reflection $R_{ss}$ at all angles.
    }}
    \label{fig3}
  \end{center}
\end{figure}

\emph{General angle of  incidence.---}
The foregoing analysis can be generalized to all incident angles. We begin by considering $a<\lambda/2$, which ensures a single diffraction order for all incident plane waves, $\mathbf{E}_{0,\mathbf{k}_{\parallel}}e^{i\mathbf{k}_{\parallel}\cdot \mathbf{r}}e^{i k_z z}$, at any angle. Here $\mathbf{k}_{\parallel} = (k_x, k_y, 0)$ denotes the projection of the incident wave vector onto the $xy$ plane and $\mathbf{E}_{0,\mathbf{k}_{\parallel}}$ can be decomposed into the two possible transverse polarizations $\mathbf{e}^+_{p,s} \bot \mathbf{k}$, see \fig{fig3}a. The total field has the form of \eq{En} where the scattering amplitude now becomes a $3\times3$ matrix and with $e^{i\mathbf{k}_\parallel \cdot \mathbf{r}_\parallel}\mathbf{E}_{0,\mathbf{k}_{\parallel}}$ replacing $\mathbf{E}_0$.
The scattering amplitude is again determined by the polarization of the atoms, which is spatially modulated by the in-plane incident wavevector, according to Bloch's theorem. The polarization of atom $n$ can thus be written as $\mathbf{p}_n=\mathbf{p}(\mathbf{k}_{\parallel})e^{i\mathbf{k}_{\parallel}\cdot\mathbf{r}_n}$, where
\begin{equation}
  \mathbf{p}(\mathbf{k}_{\parallel})=\overline{\overline{\alpha}}_e(\mathbf{k}_{\parallel})\mathbf{E}_{0,\mathbf{k}_{\parallel}}
  \label{pa}
\end{equation}
denotes the polarization in momentum space. Hence, the effective polarizability is generally defined as the linear response of the polarization of the array in momentum space, given by the tensor
\begin{equation}
  \tensor{\alpha}_e(\mathbf{k}_\parallel) = - \frac{3}{4 \pi^2} \varepsilon_0 \lambda^3
  \frac{\gamma/2}
  {\delta-\tensor{\Delta}(\vect{k}_\parallel) + i[ \gamma+\gamma_{\mathrm{nr}}+\tensor{\Gamma}(\mathbf{k}_\parallel)]/2}.
  \label{alpa}
\end{equation}
In analogy with Eqs.~\noeq{alp} and \noeq{Sn}, $\tensor{\Delta}(\vect{k}_\parallel)$ and  $\tensor{\Gamma}(\mathbf{k}_\parallel)$ are the cooperative resonance and width tensors, respectively, given in terms of the dyadic Green's function $\tensor{G}$ by
\begin{equation}
  \overline{\overline{\Delta}}(\mathbf{k}_{\parallel})-\frac{i}{2}\overline{\overline{\Gamma}}(\mathbf{k}_{\parallel}) = -\frac{3}{2}\gamma \lambda \sum_{n\neq 0} \overline{\overline{G}}(0,\mathbf{r}_n)e^{i\mathbf{k}_{\parallel}\cdot\mathbf{r}_n}.
  \label{Ga}
\end{equation}
An analytic expression can be obtained for $\tensor{\Gamma}$ \cite{SM}, while $\tensor \Delta$ has been evaluated numerically.

The scattering amplitude is related to the effective polarizability $\tensor{\alpha}_e(\mathbf{k}_\parallel)$ by an expression similar to that in \eq{Sn}, from which we can deduce the intensity reflection and transmission coefficients, $R_{\mu\nu}$ and $T_{\mu\nu}$, where $\mu$ and $\nu$ denote the polarization, either $p$ or $s$, of the reflected/transmitted and the incident field, respectively \cite{SM}. Figures~\nofig{fig3}b and \nofig{fig3}c display $R_{ss}$ and $R_{pp}$ as a function of the incident angle (represented by $k_x$, $k_y$) in the case of $a=0.2\lambda$ and $\delta=0$, for which full reflection was obtained at normal incidence. Remarkably, we observe that the reflection coefficient exceeds $0.99$ at all incident angles for $s$-polarized light. In the case of $p$-polarization, reflection exceeds $0.95$ for a wide range of angles, far beyond the paraxial approximation. Furthermore, mixing between polarizations is small $T_{sp},T_{ps},R_{sp},R_{ps} \sim 10^{-3}$ \cite{SM}, which demonstrates that the array operates as an excellent mirror for almost all incident angles and both incident polarizations.
The fact that the scattering resonance found at normal incidence persists for incident angles well beyond the paraxial regime, implies that the mirror should operate well for realistic finite size incident beams and arrays. This was verified for Gaussian beams with a waist smaller than the array size, using a direct numerical approach \cite{SM}.

The high reflection at oblique angles may again be understood in terms of cooperative resonances of the atom array. For example, in Fig. 3d we plot the $ss$ matrix element of $\tensor \Delta$, which is seen to vary by less than an atomic linewidth over all incident angles. This provides an explanation for the excellent reflection of $s$-polarized light. For $p$-polarized light, a similar explanation holds only within the paraxial regime, beyond which the discussion is complicated by the polarization degree of freedom \cite{SM}.
\begin{figure}
  \begin{center}
    \includegraphics[width=\columnwidth]{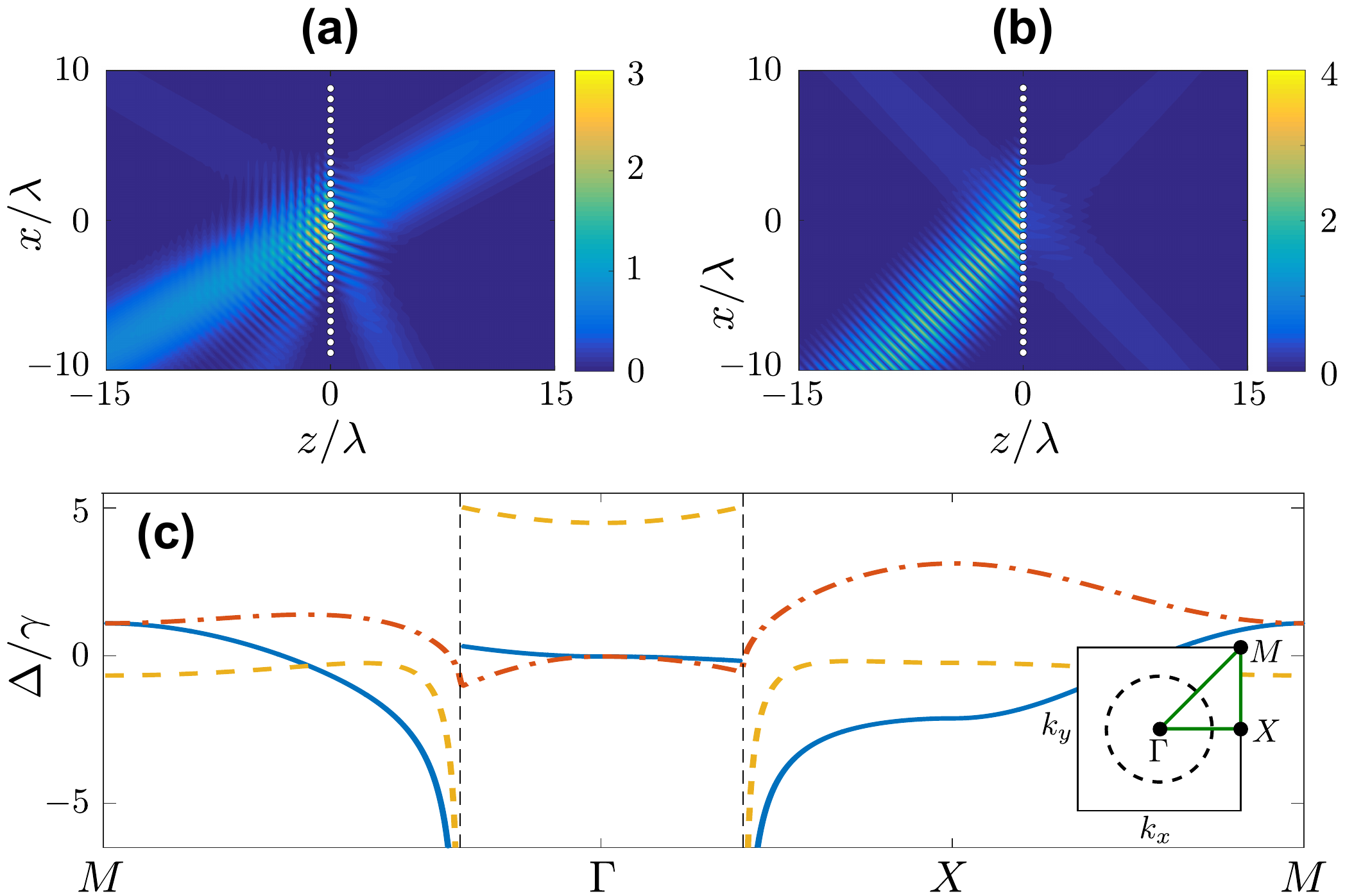}
    \caption{\small{
      (a) Scattering beyond a single diffraction order. Compared to Figs.~\nofig{fig1}b and \nofig{fig3}a,  an additional diffraction order appears on both sides of the array. The figure presents results of a numerical calculation \cite{SM} for an $s$-polarized Gaussian beam of waist $3 \lambda$ and $\delta=\Delta$, incident at a $30^\circ$ angle relative to the $z$ axis. The beam is scattered off an array of $26\times26$ atoms with $a=0.707\lambda$. All field intensities are in units of $|\mathbf{E}_0|^2$. (b) Retro-reflector effect for the same array and incident Gaussian beam, this time $p$-polarized at an incident angle of $45^\circ$. Scattering is strongly suppressed into directions for which the $p$-polarization is orthogonal to the $p$-polarization of the incident beam. As a result, most of the light is back scattered into the first diffraction order, which is parallel to the incident beam. (c) Band structure of the collective surface modes of the atom array for $a = 0.2\lambda$. The three bands correspond to the three eigenvalues of the cooperative shift $\tensor{\Delta}(\vect{k}_\parallel)$. The modes in the yellow (dashed) band are polarized along the $z$ direction, while the polarizations of the other two bands lie in the $xy$ plane. The inset shows the location of the special points $\Gamma$, $X$, $M$ in the first Brillouin zone. The dashed vertical lines (main figure) and the dashed circle (inset) indicate the light cone $|\vect{k}_\parallel| = 2 \pi / \lambda$.
    }}
    \label{fig4}
  \end{center}
\end{figure}

\emph{Beyond single diffraction order.---}
An additional diffraction order can appear when the lattice constant exceeds $\lambda/2$ (Fig. \ref{fig4}a), since the addition of a reciprocal lattice vector, $\sim 2\pi/a$, to the incident wavevector due to Bragg scattering by the lattice, may now result in a propagating wave. This situation can be analyzed by a straightforward extension of the above formalism. For example, in the case of $p$-polarization, where the propagation along certain directions is suppressed, new possibilities arise, such as the retro-reflection effect in \fig{fig4}b.

\emph{Surface dipole excitations.---}
More insight into the physics of the array is gained by noting that the cooperative shift $\overline{\overline{\Delta}}(\mathbf{k}_{\parallel})$ in fact describes the dispersion relation of collective surface dipole excitations. The nature of these surface modes is revealed by \eq{pa} as the normal modes of the atomic dipoles on the surface, $\mathbf{p}(\mathbf{k}_{\parallel})$. The resonant frequencies of the modes and their corresponding polarizations can be deduced from their linear response $\overline{\overline{\alpha}}_e(\vect{k}_\parallel)$ in \eq{alpa} as the three eigenvalues and eigenvectors of $\overline{\overline{\Delta}}(\vect{k}_{\parallel})$. This interpretation also follows from the quantum master equation governing the dynamics of the atoms \cite{SM}. In that fully equivalent treatment, the eigenvalues of $\tensor{\Delta}(\vect{k}_\parallel)$ arise naturally as the energies of the Bloch modes of atomic excitations, while $\tensor{\Gamma}(\vect{k}_\parallel)$ gives their respective decay rate.

When considering the scattering problem, $k_x$ and $k_y$ are restricted to be less than $2 \pi/\lambda$. On the other hand, the discussion of collective surface excitation requires no such assumption and we may extend $k_x$ and $k_y$ to the entire Brillouin zone $[-\pi/a,\pi/a]$. By diagonalizing $\tensor{\Delta}$ for each $\vect{k}_\parallel$ we hence obtain the band structure of the surface modes shown in \fig{fig4}c. The modes near the center of the Brillouin zone ($\Gamma$), between the vertical dashed lines, carry a crystal momentum smaller than $2 \pi/\lambda$ and may thus couple to far-field radiation. These are precisely the modes that give rise to high reflection when driven resonantly. Beyond the vertical dashed lines in \fig{fig4}c, where $|\vect{k}_\parallel| > 2 \pi/\lambda$, the surface can no longer couple to far-field radiation and these modes are thus protected against decay, satisfying $\tensor{\Gamma}+\gamma= 0$. As such, these modes can only be excited in the near field and could potentially guide light along the surface.

\emph{Experimental considerations.---}
Three promising approaches for experimental realization of the 2D array can be considered. First, an array of ultracold atoms may be trapped in an optical lattice. For the single scattering order regime, $a < \lambda/2$, a blue-detuned trapping laser can be used \cite{WEI}.
In a second implementation, the dipole array can be realized using plasmonic nano-particles~\cite{GDA1,GDA2}. Our results should readily apply to this situation, provided the collective radiative decay exceeds the individual non-radiative losses, as discussed above. Finally, 2D semiconductors such as monolayers of transition metal dichalcogenides~\cite{STR} can be used, in which excitons and trions with near lifetime limited linewidths have recently been observed~\cite{Robert2016}. A lattice structure for the excitons (trions) can be created by periodically modulating strain~\cite{ZhengLi,Atature2016}, varying the dielectric environment~\cite{Palacios2014}, or by applying an external electrostatic potential. In addition to giving rise to the unusual dispersion relation discussed above, confining the excitons (trions) to sub-wavelength scales may improve their optical properties by overcoming defect-induced localization \cite{LOC}.

Any of these realizations may also be subject to \emph{disorder} which may affect the cooperative resonances and scattering of the array. Interestingly, we show in \cite{SM} that the cooperative resonances are robust to fluctuations in the atomic positions when the fluctuations are much smaller than the lattice period. In addition, we find that for small values of  $a/\lambda$, scattering is also robust to defects such as a missing atom.

\emph{Discussion and prospects.---}
The above considerations demonstrate that the scattering properties of light at any incident polarization and detuning $\delta$ off a 2D atomic array can be controlled by choosing the lattice spacing of the array $a$, enabling, for instance, its operation as a nearly perfect mirror.

These results suggest the potential use of such 2D arrays as powerful platforms for classical and quantum optics.
While our results are obtained for a uniform atomic array, it should be possible to generalize our approach to non-homogeneous arrays, resulting in the realization of ``atomic-scale  metasurfaces'' with desired properties.  One particularly intriguing possibility is to engineer the surfaces such that a given incident mode is effectively focused on a single impurity atom. While even for the uniform array we find optical nonlinearities at the level of $\sim10$ photons \cite{SM}, such engineered arrays will likely display nonlinearities down to a single photon level similar to 1D systems \cite{CHA}.
Alternatively, exceptional  nonlinearities can be realized by combing the present approach with Rydberg blockade and electromagnetically-induced transparency \cite{LUKR,OFERR,PFAU}.

This work also opens up  new prospects in the field of optomechanics. Since the atoms are very light but at the same time collectively exhibit nearly perfect reflection, they form a highly mechanically-susceptible mirror, potentially very useful for the exploration of optomechanics at the quantum level \cite{OPM}.

Finally, we stress the universality of our approach for any physical system of waves and dipole-like scatterers. In particular, our analysis of cooperative resonances is not limited to electric dipoles, and the electromagnetic Green's function can be readily replaced by a Green's function describing an entirely different wave phenomenon at any wavelength.

\begin{acknowledgements}
We thank J\'{a}nos Perczel for insightful comments concerning the quantum description and dispersion relation of the array. We also acknowledge valuable discussions with Vladimir Shalaev, Markus Greiner, Peter Zoller, Darrick Chang, Hongkun Park, Alex High and Kristiaan de Greve, and financial support from NSF and the MIT-Harvard Center for Ultracold Atoms.
\end{acknowledgements}

\pagebreak
\widetext
\begin{center}
\textbf{\large Supplemental Materials: Cooperative resonances in light scattering from two-dimensional atomic arrays}
\end{center}
\setcounter{equation}{0}
\setcounter{figure}{0}
\setcounter{table}{0}
\setcounter{page}{1}
\makeatletter
\renewcommand{\theequation}{S\arabic{equation}}
\renewcommand{\thefigure}{S\arabic{figure}}
\renewcommand{\bibnumfmt}[1]{[S#1]}
\renewcommand{\citenumfont}[1]{S#1}

This Supplementary document is organized as follows. Sec. 1 reviews the general approach we use for the scattering problem by an array of dipole scatterers. Sec. 2 is dedicated for the derivation of our analytical formulation of scattering by a 2D lattice, including its quantum version. Sec. 3-5 discuss in more depth some of the results and consequences that emerge from the our analytical treatment. Finally, Sec. 6 presents the direct numerical approach we employ to verify our theoretical results in a realistic finite size scenario.

\section{1. The scattering problem: General approach}
\label{SM_sec_gen}
We consider the scattering of a general incident electric field by a collection of atoms modeled as point-like dipole scatterers characterized by their linear polarizability. This description naturally applies for the scattering due to the electric linear response of any collection of scatterers which are much smaller than the incident wavelength, such as atoms in the optical frequency domain, and has been used before by many authors in the photonics community, e.g. for the treatment of light scattering by arrays of nano-particles \cite{sGDA1}.

\subsection{1.1 Electromagnetic scattering theory}
Assuming an incident field $\mathbf{E}_0(\mathbf{r})$ and atoms $n=1,..,N$ at positions $\mathbf{r}_n$, the goal is to find the electric field at any given point $\mathbf{r}$. As usual, we begin with Maxwell's curl equations for fields at frequency $\omega=kc=k/\sqrt{\varepsilon_0\mu_0}$, obtaining the wave equation
\begin{equation}
\nabla\times\nabla\times\mathbf{E}-k^2\mathbf{E}=\frac{k^2}{\varepsilon_0}\mathbf{P},
\label{SM_helm}
\end{equation}
whose formal solution is
\begin{equation}
E_i(\mathbf{r})=E_{0,i}(\mathbf{r})+\frac{k^2}{\varepsilon_0}\sum_j\int_V d\mathbf{r}'G_{ij}(k,\mathbf{r},\mathbf{r}')P_j(\mathbf{r}'),
\label{SM_sol}
\end{equation}
with $A_i=\mathbf{e}_i\cdot\mathbf{A}$ the $i=x,y,z$ component of a vector $\mathbf{A}$ and where $G_{ij}(k,\mathbf{r},\mathbf{r}')$ is the Green's function of Eq. (\ref{SM_helm}), namely the $i$ component of electric field (in inverse length units) given a delta-function source at $\mathbf{r}'$ polarized on the $j$ axis; in free space this so-called dyadic Green's function takes the form \cite{sNH}
\begin{equation}
G_{ij}(k,\mathbf{r}_1,\mathbf{r}_2)=\frac{e^{ikr}}{4\pi r}\left[\left(1+\frac{ikr-1}{k^2r^2}\right)\delta_{ij}+\left(-1+\frac{3-3ikr}{k^2r^2}\right)\frac{r^ir^j}{r^2}\right],
\label{SM_G}
\end{equation}
with $\mathbf{r}=\mathbf{r}_1-\mathbf{r}_2$, $r=|\mathbf{r}|$ and $r^i=\mathbf{e}_i\cdot\mathbf{r}$.

Considering now that polarization $\mathbf{P}$ exists only on the atoms and using their linear response $\alpha(\omega)$ (generally a tensor, but taken as scalar/isotropic and identical for all atoms), we have
\begin{equation}
\mathbf{P}(\mathbf{r})=\sum_{n=1}^N\alpha  \mathbf{E}(\mathbf{r})\delta(\mathbf{r}-\mathbf{r}_n),
\label{SM_Pol}
\end{equation}
so that Eq. (\ref{SM_sol}) becomes
\begin{equation}
E_i(\mathbf{r})=E_{0,i}(\mathbf{r})+4\pi^2\frac{\alpha}{\varepsilon_0\lambda^3} \sum_j\sum_n \lambda G_{ij}(k,\mathbf{r},\mathbf{r}_n)E_j(\mathbf{r}_n),
\label{SM_LS}
\end{equation}
where $\lambda=2\pi/k$.
Eq. (\ref{SM_LS}) has a Lippman-Schwinger form, and together with the Green's function, Eq. (\ref{SM_G}), it forms the formal solution and the starting point of our scattering theory. The idea is to self-consistently evaluate the fields $E_j(\mathbf{r}_n)$, or more precisely the polarizations $p_j(\mathbf{r}_n)=\alpha E_j(\mathbf{r}_n)$, at the atomic positions $\mathbf{r}=\mathbf{r}_n$, which in turn determine the right-hand side of the equation, thus allowing for solution of the field at any point, $E_i(\mathbf{r})$ at the left-hand side of the equation. Writing Eq. (\ref{SM_LS}) at an atomic position $\mathbf{r}=\mathbf{r}_n$, we obtain a self-consistent equation for the polarization induced on the atoms $p^n_i= \alpha E_i(\mathbf{r}_n)$,
\begin{equation}
p^n_i=p^n_{0,i}+\sum_{m\neq n} \sum_j 4\pi^2\frac{\alpha}{\varepsilon_0\lambda^3} \lambda G^{nm}_{ij}p^m_j,
\label{SM_LSp}
\end{equation}
with $p^n_{0,i}=\alpha E_{0,i}(\mathbf{r}_n)$ and $G^{nm}_{ij}=G_{ij}(k,\mathbf{r}_n,\mathbf{r}_m)$. Here the summation over atoms (index $m$) skips the atom $n$ at which we evaluate the field, in accordance with the convention we adopt for the atomic polarizability, as explained below. The idea is that by inserting the solution of Eq. (\ref{SM_LSp}) into Eq. (\ref{SM_LS}), we obtain the field at any point $E_i(\mathbf{r})$.

The solution of Eq. (\ref{SM_LSp}) and then of Eq. (\ref{SM_LS}) can be readily performed numerically for any general case of incident field and finite collection of atoms, as we show in Sec. \ref{SM_sec_num} below. Nevertheless, by exploiting the symmetry of an ordered array, an analytical solution can be obtained, as shown in Sec. \ref{SM_sec_th} (similar to the approach taken in Ref. \cite{sGDA1}).

\subsection{1.2 The atomic polarizability}
Considering a $J=0$ to $J=1$ transition of an atom with a dipole matrix element $d$ and frequency $\omega_a$ the linear polarizability is given by \cite{sMQED,sPET}
\begin{equation}
\alpha(\omega)=\frac{2d^2\omega_a}{\hbar}\frac{1}{\omega_a^2-\omega^2-i\gamma_0\omega},
\label{SM_alp1}
\end{equation}
where $\gamma_0$ is a damping term which is addressed below. It is worth recalling here that a similar form of polaizability is found for classical systems alike, so that our discussion is kept general for all point-like polarizable scatterers in their linear regime. For small detuning $\delta=\omega-\omega_a$ with respect to $\omega_a$, $|\delta|\ll\omega_a$, and using the spontaneous emission rate of a single atom $\gamma=d^2\omega_a^3/(3\pi\varepsilon_0\hbar c^3)$, we obtain
\begin{equation}
\alpha(\omega)=-\frac{3}{4\pi^2}\varepsilon_0\lambda_a^3\frac{\gamma/2}{\delta+i\gamma_0/2}.
\label{SM_alp}
\end{equation}
Skipping the $m=n$ term in the sum of Eq. (\ref{SM_LSp}) implies that the radiative damping $\gamma$ of an individual atom is included in the total damping $\gamma_0$. Namely,
\begin{equation}
\gamma_0=\gamma+\gamma_{\mathrm{nr}},
\label{SM_gam0}
\end{equation}
where $\gamma_{\mathrm{nr}}$ denotes the non-radiative loss rate. This also implies that $\omega_a$ already includes the Lamb shift correction (or that we neglect it). We note, that one may chose to include the $m=n$ term in Eq. (\ref{SM_LSp}) by neglecting the real part of $G^{nn}_{ij}$ (neglecting the Lamb shift) and taking $\gamma_0=\gamma_{\mathrm{nr}}$, yielding exactly the same results for $p_i^n$ and $E_i(\mathbf{r})$.

\section{2. Analytical approach}
\label{SM_sec_th}
\subsection{2.1 Polarization built on the atoms}
We consider an infinite array of atoms spanning the $xy$ plane at $z=0$ and illuminated by an incident field represented by its decomposition to plane waves $\mathbf{E}_0(\mathbf{r})=\sum_{\mathbf{k}_{\parallel}}\mathbf{E}_{0,\mathbf{k}_{\parallel}} e^{i\mathbf{k}\cdot \mathbf{r}}$. The wavevector of each plane wave is characterized by its projections on the $xy$ plane  and $z$ axis, $\mathbf{k}_{\parallel}$ and $k_z=\sqrt{k^2-|\mathbf{k}_{\parallel}|^2}$, respectively, whereas its amplitude $\mathbf{E}_{0,\mathbf{k}_{\parallel}}$ is spanned by the two transverse polarizations $\mathbf{e}^+_{p,s} \bot \mathbf{k}$. At normal incidence ($\mathbf{k}_{\parallel}=0$) we have $k_z=k$ and the field is polarized in the $xy$ plane, $\mathbf{e}^+_{p,s}\in\{\mathbf{e}_x,\mathbf{e}_y\}$.

In order to exploit the discrete translational symmetry of the lattice (as in the Bloch theorem), we define the 2D Fourier transform of a function $f(\mathbf{r})$ sampled at the lattice sites $n$, $f_n=f(\mathbf{r}_n)$, as
\begin{equation}
f(\mathbf{k}_{\parallel})=\frac{1}{N}\sum_{n=1}^Ne^{-i\mathbf{k}_{\parallel}\cdot\mathbf{r}_n}f_n,
\quad
f_n=N A_0\int_{\mathrm{BZ}} \frac{d\mathbf{k}_{\parallel}}{(2\pi)^2}e^{i\mathbf{k}_{\parallel}\cdot\mathbf{r}_n}f(\mathbf{k}_{\parallel}),
\label{SM_FT}
\end{equation}
where $A_0$ is the area of the lattice unit cell and BZ means that the integral over $\mathbf{k}_{\parallel}$ is performed within the first Brillouin zone of the reciprocal lattice. For a square lattice, $\mathbf{r}_n=\mathbf{r}_{n_x,n_y}=a(n_x\mathbf{e}_x+n_y\mathbf{e}_y)$, we have $A_0=a^2$ and $\int_{\mathrm{BZ}} d\mathbf{k}_{\parallel}=\int_{-\pi/a}^{\pi/a}dk_x\int_{-\pi/a}^{\pi/a}dk_y$.
Applying this Fourier transformation on Eq. (\ref{SM_LSp}) we find
\begin{equation}
\mathbf{p}(\mathbf{k}_{\parallel})=\alpha \mathbf{E}_{0,\mathbf{k}_{\parallel}}+4\pi^2\frac{\alpha}{\varepsilon_0\lambda^3} \lambda \overline{\overline{g}}(\mathbf{k}_{\parallel})\mathbf{p}(\mathbf{k}_{\parallel})
\label{SM_LSFT}
\end{equation}
for each $\mathbf{k}_{\parallel}$ value contained in the incident field, where
\begin{equation}
\overline{\overline{g}}(\mathbf{k}_{\parallel})=\sum_{n\neq 0}\overline{\overline{G}}(0,\mathbf{r}_n)e^{-i\mathbf{k}_{\parallel}\cdot\mathbf{r}_n},
\label{SM_g}
\end{equation}
also noting the fact that the dyadic Green's function in free space, $\overline{\overline{G}}(\mathbf{r},\mathbf{r}')$ depends only on the difference $\mathbf{r}-\mathbf{r}'$, see Eq. (\ref{SM_G}). The solution of (\ref{SM_LSFT}) is then
\begin{eqnarray}
&&\mathbf{p}_n=\sum_{\mathbf{k}_{\parallel}}\mathbf{p}(\mathbf{k}_{\parallel})e^{i\mathbf{k}\cdot \mathbf{r}_n},
\nonumber \\
&&\mathbf{p}(\mathbf{k}_{\parallel})=\overline{\overline{\alpha}}_e(\mathbf{k}_{\parallel})\mathbf{E}_{0,\mathbf{k}_{\parallel}},
\label{SM_p}
\end{eqnarray}
with the effective polarizability tensor
\begin{equation}
  \tensor{\alpha}_e(\mathbf{k}_\parallel) = - \frac{3}{4 \pi^2} \varepsilon_0 \lambda_a^3
  \frac{\gamma/2}
  {\delta-(\lambda_a/\lambda)^3\tensor{\Delta}(\vect{k}_\parallel) + i[ \gamma+\gamma_{\mathrm{nr}}+(\lambda_a/\lambda)^3\tensor{\Gamma}(\mathbf{k}_\parallel)]/2},
    \label{SM_alpe}
\end{equation}
and the collective radiative response from Eq. (7) of the main text, namely,
\begin{equation}
\tensor{\Delta}(\vect{k}_\parallel)=-\frac{3}{2}\gamma\lambda\mathrm{Re}[\tensor{g}(\vect{k}_\parallel)], \quad \tensor{\Gamma}(\vect{k}_\parallel)=3\gamma\lambda\mathrm{Im}[\tensor{g}(\vect{k}_\parallel)].
  \label{SM_DG}
\end{equation}
Finally, the effective polarizability from Eq. (6)  of the main text is reached by taking $\lambda_a^3/\lambda^3\approx1$ in Eq. (\ref{SM_alpe}), equivalent to the Markov approximation $\omega^3\approx\omega_a^3$ in quantum optics. We note that Eq. (\ref{SM_p}) implies that the polarization built on the atomic array due to an incident plane wave with in-plane wavevector $\mathbf{k}_{\parallel}$ is not mixed with that induced by a different wavevector, and that the \emph{collective} response of the array polarization to a field with a given wavenumber $\mathbf{k}_{\parallel}$ is described by the effective polarizability $\overline{\overline{\alpha}}_e(\mathbf{k}_{\parallel})$.

The result at normal incidence, Eq. (3) in the main text, is obtained by noting that the sum $\tensor{g}(0)$ for $\mathbf{k}_{\parallel}=0$ in Eq. (\ref{SM_g}) is symmetric in the $x$ and $y$ directions due to the structure of $\tensor{G}$ from Eq. (\ref{SM_G}) and the symmetry of the lattice. Therefore, the tensor $\tensor{g}$ becomes diagonal $g_{ij}=\delta_{ij}g_i$ with $g_x=g_y$. At normal incidence the field is polarized in the $xy$ plane so that $\tensor{g}$ effectively appears as a scalar, $g_{x}=\sum_{n\neq 0}G_{xx}(0,\mathbf{r}_n)$, and the collective response becomes $\Delta-i\Gamma/2=-(3/2)\gamma\lambda g_x$.

\subsection{2.2 Calculation of the cooperative shift and width}
Both $\tensor{\Delta}$ and $\tensor{\Gamma}$ are obtained by a direct numerical summation over the atomic lattice points, as in Eq. (\ref{SM_DG}). Alternatively, one can begin by representing the sum $\tensor{g}(\vect{k}_\parallel)$ in 2D reciprocal (wavevector) space by using the relation \cite{sNH}
\begin{equation}
G_0(x,y,z)=\frac{e^{ik\sqrt{x^2+y^2+z^2}}}{4\pi\sqrt{x^2+y^2+z^2}}=\frac{i}{8\pi^2}\int d\mathbf{k}'_{\parallel}e^{-i\mathbf{k}'_{\parallel}\cdot \mathbf{r}_{\parallel}}\frac{e^{ik'_z |z|}}{k'_z},
\label{SM_G0}
\end{equation}
where $\mathbf{r}_{\parallel}=(x,y)$ and $k'_z=\sqrt{k^2-|\mathbf{k}'_{\parallel}|^2}$. Then, writing
\begin{equation}
g_{ij}(\vect{k}_\parallel)=\sum_{n}\int d\mathbf{r}_{\parallel} G_{ij}(0,\mathbf{r}_{\parallel})e^{-i\mathbf{k}_{\parallel}\cdot\mathbf{r}_{\parallel}}\delta(\mathbf{r}_{\parallel}-\mathbf{r}_n)-G_{ij}(0,0)
\label{SM_ggg}
\end{equation}
(recalling that $\mathbf{r}_n$ only exists on the $xy$ plane), and using $G_{ij}=[\delta_{ij}+(1/k^2)\partial_i\partial_j]G_0$, we find
\begin{equation}
g_{ij}(\vect{k}_\parallel)=
\left\{
  \begin{array}{ll}
    \frac{i}{8\pi^2}\int d\mathbf{k}'_{\parallel}\frac{1}{k'_z}\left(\delta_{ij}-\frac{k'_i k'_j}{k^2}\right)\sum_n\int\mathbf{r}_{\parallel}\delta(\mathbf{r}_{\parallel}-\mathbf{r}_n)e^{-i(\mathbf{k}'_{\parallel}+\mathbf{k}_{\parallel})\cdot\mathbf{r}_{\parallel}}-\frac{i}{3\lambda}\delta_{ij}, \quad \mathrm{for} \quad i,j=\{x,y\} \cup i=j=z,
    \\
    -\frac{i}{3\lambda}\delta_{ij},  \quad \mathrm{otherwise}.
  \end{array}
\right.
\label{SM_g1}
\end{equation}
Here we used $G_{ij}(0,0)=\frac{i}{3\lambda}\delta_{ij}$ obtained by taking the $\mathbf{r}\rightarrow 0$ limit of $G_{ij}(\mathbf{r},0)$ and neglecting its real part (associated with the Lamb shift of the atomic resonance $\omega_a$). We may now introduce the reciprocal lattice as the Fourier transform of the atomic lattice structure
\begin{equation}
\rho(\mathbf{k}'_{\parallel})=\int d\mathbf{r}_{\parallel} e^{-i\mathbf{k}'_{\parallel}\cdot \mathbf{r}_{\parallel}}\sum_n \delta(\mathbf{r}_{\parallel}-\mathbf{r}_n)=\frac{(2\pi)^2}{A_0}\sum_m \delta(\mathbf{k}'_{\parallel}-\mathbf{q}_m),
\label{SM_RL}
\end{equation}
which forms a lattice in 2D wavevector space at lattice points $\mathbf{q}_m$ (wavevectors) satisfying $\mathbf{q}_m\cdot\mathbf{r}_n=2\pi M$ with an integer $M$ and for any $n$ and $m$. Using Eq. (\ref{SM_RL}) in (\ref{SM_g1}) we obtain
\begin{equation}
g_{ij}(\vect{k}_\parallel)=\frac{i}{2A_0}\sum_m \frac{1}{\sqrt{k^2-|\mathbf{q}_m-\mathbf{k}_{\parallel}|^2}}\left[\delta_{ij}-\frac{(\mathbf{q}_m-\mathbf{k}_{\parallel})_i(\mathbf{q}_m-\mathbf{k}_{\parallel})_j}{k^2}\right]-\delta_{ij}\frac{i}{3\lambda},
\label{SM_g2}
\end{equation}
valid for $ i,j=\{x,y\}$ or $i=j=z$ whereas $g_{ij}(\vect{k}_\parallel)=-\delta_{ij}\frac{i}{3\lambda}$ otherwise. For the square lattice we have $\mathbf{q}_{m_x,m_y}=(2\pi/a)(m_x\mathbf{e}_x+m_x\mathbf{e}_y)$ and the unit cell area is $A_0=a^2$ so that Eq. (\ref{SM_g2}) becomes
\begin{equation}
g_{ij}(\vect{k}_\parallel)=\frac{i}{2a^2}\sum_{m_x=-\infty}^{\infty} \sum_{m_y=-\infty}^{\infty} \frac{\left(\delta_{ij}-[(2\pi/a)m_x-k_x][(2\pi/a)m_y-k_y]/k^2\right)}{\sqrt{k^2-[(2\pi/a)m_x-k_x]^2-[(2\pi/a)m_y-k_y]^2}}-\delta_{ij}\frac{i}{3\lambda}.
\label{SM_g3}
\end{equation}
The sums in Eqs. (\ref{SM_g2},\ref{SM_g3}) can be used for a numerical evaluation of $\tensor{g}$ (and hence of $\tensor{\Delta}$ and $\tensor{\Gamma}$) in momentum space, in contrast to the sum (\ref{SM_g}) that is performed in real space. This is achieved by using a proper regularization and renormalization procedure \cite{noteLS}. More importantly, this form of $\tensor{g}$ can be used to find an analytical expression for $\tensor{\Gamma}$ in the single diffraction order case. Considering only the imaginary part of $g_{ij}$ in Eq. (\ref{SM_g3}), the square root in the denominator has to be real. Since this square root originates in the wavenumber in the $z$ direction, $k'_z$ that appears in the wave expansion of the Green's function (Eqs. \ref{SM_G0} and \ref{SM_g1}), this requirement means that radiation damping (imaginary part of $g_{ij}$) is related to waves that can propagate out of the atomic array. Taking for example $k_y=0$, we have $|k_x-m_x 2\pi/a|<k$ and $|k_x|<k$, leading to $|m_x|<2a/\lambda$ (for $k_y\neq0$ this condition on $m_x$ becomes even more restrictive). Then, for the case $a<\lambda/2$ and for any $(k_x,k_y)$ only the zeroth-order diffraction $m_x=m_y=0$ contributes to radiation and damping. At normal incidence $k_x=k_y=0$, such single diffraction order exists even for $a<\lambda$ as can be inferred from the condition on $m_x$.
Similar conclusions are reached for any lattice structure with a scale $a$ of its unit cell by considering Eq. (\ref{SM_g2}). Then, from Eq. (\ref{SM_DG}) we obtain
\begin{equation}
\Gamma_{ij}=\gamma\frac{3}{4\pi}\left(\frac{\lambda}{a}\right)^2\frac{k}{k_z}\left(\delta_{ij}-\frac{k_i k_j}{k^2}\right)-\gamma\delta_{ij},
\label{SM_Ga}
\end{equation}
again  valid for $i,j\in\{x,y\}$ or $i=j=z$, while $\Gamma_{ij}=-\gamma\delta_{ij}$ otherwise. At normal incidence we have $k_x=k_y=0$ and $k_z=k$ leading to $\Gamma=\Gamma_{xx}=\gamma\frac{3}{4\pi}\frac{\lambda^2}{a^2}-\gamma$ as in Eq. (4) of the main text.
For larger values of $a$ more diffraction orders may appear and analytical results for their contribution to $\tensor{\Gamma}$ are obtained in the same way.

\subsection{2.3 The scattered field}
Let us now find the scattered field due to a given plane wave component of the incident field $\mathbf{E}_{0,\mathbf{k}_{\parallel}} e^{i\mathbf{k}\cdot \mathbf{r}}$ with an in-plane wavevector $\mathbf{k}_{\parallel}$. Returning to Eq. (\ref{SM_LS}), we use $\mathbf{E}^n=\mathbf{p}^n/\alpha=\overline{\overline{\alpha}}_e(\mathbf{k}_{\parallel})\mathbf{E}_{0,\mathbf{k}_{\parallel}}e^{i\mathbf{k}_{\parallel}\cdot \mathbf{r}_n}/\alpha$ in the right-hand side, and obtain
\begin{equation}
\mathbf{E}(\mathbf{r})=\mathbf{E}_{0,\mathbf{k}_{\parallel}} e^{i\mathbf{k}_{\parallel}\cdot \mathbf{r}}e^{ik_z z}+4\pi^2 \lambda \tensor{g}_{sc}(\mathbf{k}_{\parallel},\mathbf{r}) \frac{\overline{\overline{\alpha}}_e(\mathbf{k}_{\parallel})}{\varepsilon_0\lambda^3}\mathbf{E}_{0,\mathbf{k}_{\parallel}},
\label{SM_LSsc}
\end{equation}
with
\begin{equation}
\tensor{g}_{sc}(\mathbf{k}_{\parallel},\mathbf{r})=\sum_n \lambda \tensor{G}(\mathbf{r},\mathbf{r}_n)e^{i\mathbf{k}_{\parallel}\cdot \mathbf{r}_n}.
\label{SM_gsc}
\end{equation}
Using again the above methods [the expansion (\ref{SM_G0}) and the reciprocal lattice (\ref{SM_RL})] we find
\begin{equation}
g_{sc,ij}(\mathbf{k}_{\parallel},\mathbf{r})=\frac{i}{2A_0}\sum_m \frac{1}{\sqrt{k^2-|\mathbf{k}_{\parallel}+\mathbf{q}_m|^2}}\left[\delta_{ij}-\xi_{ij}\frac{(\mathbf{k}_{\parallel}+\mathbf{q}_m)_i(\mathbf{k}_{\parallel}+\mathbf{q}_m)_j}{k^2}\right]
e^{i(\mathbf{k}_{\parallel}+\mathbf{q}_m)\cdot \mathbf{r}_{\parallel}}e^{i\sqrt{k^2-|\mathbf{k}_{\parallel}+\mathbf{q}_m|^2}|z|},
\label{SM_gsc1}
\end{equation}
with $\mathbf{r}_{\parallel}$ and $z$ the projections of $\mathbf{r}$ along the $xy$ plane and the $z$ axis, respectively, and where $\xi_{ij}=-1$ for either $i=z\cap j\neq z$ or $j=z\cap i\neq z$ at $z<0$, and $\xi_{ij}=1$ otherwise. For a square lattice we have as usual $A_0=a^2$, $\sum_m\rightarrow \sum_{mx=-\infty}^{\infty}\sum_{my=-\infty}^{\infty}$ and $\mathbf{q}_m\rightarrow(2\pi/a)(m_x\mathbf{e}_x+m_y\mathbf{e}_y)$. Therefore, the field scattered from all atoms to a point $\mathbf{r}$, represented by the sum $\tensor{g}_{sc}(\mathbf{k}_{\parallel},\mathbf{r})$ from Eq. (\ref{SM_gsc}), is revealed by Eq. (\ref{SM_gsc1}) to be a sum of plane waves contributions from all diffraction orders $m$ with wavevectors $\mathbf{k}_{\parallel}+\mathbf{q}_m$ and $\sqrt{k^2-|\mathbf{k}_{\parallel}+\mathbf{q}_m|^2}$ projected along the $xy$ plane and $z$ axis, respectively.

Since we are mainly interested in fields that are scattered and which propagate away from the surface along the $\pm z$ directions, the $z$-projected wavenumber $\sqrt{k^2-|\mathbf{k}_{\parallel}+\mathbf{q}_m|^2}$ has to be real (as opposed to the case of surface modes which are evanescent along $z$). This leads to the same conditions on the diffraction orders $m$ as analyzed in the analytical calculation of $\tensor{\Gamma}$, so that for the single diffraction order case, e.g. $a<\lambda/2$ in  the square lattice case, we obtain
\begin{equation}
\lambda g_{sc,ij}(\mathbf{k}_{\parallel},\mathbf{r})=\frac{i}{4\pi}\left(\frac{\lambda}{a}\right)^2\frac{k}{k_z}\left(\delta_{ij}-\xi_{ij}\frac{k_i k_j}{k^2}\right)e^{i\mathbf{k}_{\parallel}\cdot \mathbf{r}_{\parallel}}e^{ik_z|z|}.
\label{SM_gsc2}
\end{equation}
The total field at $\mathbf{r}=(\mathbf{r}_{\parallel},z)$ with $|z|\gtrsim a$ sufficiently far from the atomic array where the evanescent waves do not contribute, is then found by inserting the above $\tensor{g}_{sc}(\mathbf{k}_{\parallel},\mathbf{r})$ into Eq. (\ref{SM_LSsc}), yielding
\begin{equation}
\vect{E} (\mathbf{r})= \left[  e^{i k_z z}+ \tensor{S}_\pm (\mathbf{k}_\parallel) \, e^{ik_z |z|}\right] e^{i\mathbf{k}_\parallel \cdot \mathbf{r}_\parallel}\mathbf{E}_{0,\mathbf{k}_{\parallel}} ,
\label{SM_Ea}
\end{equation}
with the scattering matrix
\begin{equation}
\tensor{S}_\pm (\mathbf{k}_\parallel)=\frac{4\pi^2}{\epsilon_0 \lambda^2}e^{-i\mathbf{k}_{\parallel}\cdot \mathbf{r}_{\parallel}}e^{-ik_z|z|}\tensor{g}_{sc}(\mathbf{k}_{\parallel},\mathbf{r}) \tensor{\alpha}_e(\mathbf{k}_\parallel).
\label{SM_S1}
\end{equation}
The subscript $\pm$ in $\tensor{S}_\pm$ distinguishes between forward ($+$ sign for $z>0$) and backward ($-$ sign for $z<0$) scattering. The fact that $\tensor{S}_\pm$ depends on the sign of $z$, without $z$ explicitly appearing in it, is due to its dependence on $\xi_{ij}$ which is sensitive to the sign of $z$ (see explanations below Eq. \ref{SM_gsc1}).

It is constructive to analyze the scattering within the field polarization basis which is most natural for propagating plane waves, namely, that of the wavevector $\mathbf{k}=k\mathbf{e}_k$ and the transverse polarizations perpendicular to it. Eq. (\ref{SM_Ea}) reveals that for an incoming plane wave with a wavevector $\mathbf{k}=(\mathbf{k}_{\parallel},k_z)$, an additional wavevector  $\mathbf{k}'=(\mathbf{k}_{\parallel},-k_z)$ emerges for the backward scattered field (at $z<0$) while the forward scattered field (at $z>0$) possess the incident wavenumber $\mathbf{k}$. We can therefore define two relevant polarization basis, one for the incoming and transmitted fields and the other for reflected ones,  $\{\mathbf{e}_k,\mathbf{e}_p^+,\mathbf{e}_s^+\}$ and $\{\mathbf{e}_{k'},\mathbf{e}_p^-,\mathbf{e}_s^-\}$, respectively, with $\mathbf{e}_{p,s}^+\bot\mathbf{e}_k$ and $\mathbf{e}_{p,s}^-\bot\mathbf{e}_{k'}$ (Fig. 1b in the main text). In spherical coordinates these basis vectors read
\begin{eqnarray}
  &&\mathbf{e}_k=(\sin\theta\cos\phi, \sin\theta\sin\phi,\cos\theta), \quad
  \mathbf{e}_p^+=(\cos\theta\cos\phi, \cos\theta\sin\phi,-\sin\theta), \quad
  \mathbf{e}_s^+=(\sin\phi, -\cos\phi,0);
  \nonumber\\
  &&\mathbf{e}_{k'}=(\sin\theta\cos\phi, \sin\theta\sin\phi,-\cos\theta), \quad
  \mathbf{e}_p^-=(-\cos\theta\cos\phi, -\cos\theta\sin\phi,-\sin\theta), \quad
  \mathbf{e}_s^-=(\sin\phi, -\cos\phi,0),
\label{SM_pol2}
\end{eqnarray}
where all column vectors here are written in cartesian basis, i.e. as $(A_x,A_y,A_z)$ for a vector $\mathbf{A}$.

Writing $\tensor{g}_{sc}$ from Eq. (\ref{SM_gsc2}) as
\begin{equation}
\lambda\tensor{g}_{sc}(\mathbf{k}_{\parallel},\mathbf{r})=\frac{i}{4\pi}\left(\frac{\lambda}{a}\right)^2\frac{k}{k_z}\tensor{F}_{\pm}e^{i\mathbf{k}_{\parallel}\cdot\mathbf{r}_{\parallel}}e^{ik_z|z|},
\label{SM_gsc3}
\end{equation}
with
\begin{equation}
F_{\pm,ij}=\delta_{ij}-\xi_{ij}k_i k_j/k^2
\label{SM_F}
\end{equation}
being its tensor part (written in cartesian basis) and where the $\pm$ subscript is due to the dependence of $\xi_{ij}$ on the sign of $z$, we notice that the $\{\mathbf{e}_k,\mathbf{e}_p^+,\mathbf{e}_s^+\}$ and $\{\mathbf{e}_{k'},\mathbf{e}_p^-,\mathbf{e}_s^-\}$ basis form the eigenvectors of $\tensor{F}_+$ and $\tensor{F}_-$, respectively, with eigenvalues $\{0,1,1\}$, respectively, for both $\pm$ cases. Considering also that the incident field must posses one of two forward transverse polarizations $\mathbf{e}^+_{p,s}$, we can now describe the total field in Eq. (\ref{SM_Ea}) via its components $E^{\mu}=\mathbf{e}_{\mu}\cdot \mathbf{E}^{\mu}$ projected onto the transverse polarization basis $\mu=\{p,s\}$,
\begin{equation}
E^{\mu}(\mathbf{r})=\sum_{\nu=p,s}\left[\delta_{\mu\nu}e^{i k_z z}+S^{\pm}_{\mu\nu}(\mathbf{k}_{\parallel})e^{ik_z |z|}\right]e^{i\mathbf{k}_{\parallel}\cdot \mathbf{r}_{\parallel}} E^{\nu}_{0,\mathbf{k}_{\parallel}} .
\end{equation}
Here the scattering matrix is represented by its $2\times 2$ matrix elements, $S_{\mu\nu}^{\pm}=\mathbf{e}^{\pm\dag}_{\mu}\tensor{S}\mathbf{e}^+_{\nu}$, which are obtained form Eqs. (\ref{SM_S1},\ref{SM_gsc3}) and the matrix elements $\mathbf{e}^{\pm\dag}_{\mu}\tensor{F}_{\pm}\mathbf{e}^{\pm}_{\nu}=\delta_{\mu\nu}$ as
\begin{equation}
S_{\mu\nu}^{\pm}(\mathbf{k}_{\parallel})=i\pi\left(\frac{\lambda}{a}\right)^2\frac{k}{k_z}\frac{1}{\varepsilon_0\lambda^3} \mathbf{e}^{\pm\dag}_{\mu}\overline{\overline{\alpha}}_e(\mathbf{k}_{\parallel})\mathbf{e}^+_{\nu}.
\label{SM_Sa}
\end{equation}
The scattering properties of the array can then be further characterized by the intensity transmission and reflection matrices, $T_{\mu\nu}=|\delta_{\mu\nu}+S^+_{\mu\nu}|^2$ and $R_{\mu\nu}=|S^-_{\mu\nu}|^2$, where $\delta_{\mu\nu}$ is the Kronecker delta.

At normal incidence we recall that $\overline{\overline{\alpha}}_e$ is a scalar in the $\{x,y\}$ basis (which forms the $\{p,s\}$-polarization basis). Then, $\tensor{S}$ is also a scalar $S_{\mu\nu}=S\delta_{\mu\nu}$ where $S$ is given by Eq. (\ref{SM_Sa}) with $k_z=k$ and with the scalar polarizability at normal incidence (see text below Eq. \ref{SM_alpe}),
\begin{equation}
\alpha_e = - \frac{3}{4 \pi^2} \varepsilon_0 \lambda_a^3
  \frac{\gamma/2}
  {\delta-\Delta + i( \gamma+\gamma_{\mathrm{nr}}+\Gamma)/2},
    \label{SM_alpen}
\end{equation}
equivalent to Eq. (3) in the main text.

\subsection{2.4 Quantum master equation formulation}
We now extend our discussion to treat both the atoms and the electromagnetic field quantum mechanically. An isotropic atom must consist of at least four levels: a ground state $\ket{g}$ and three degenerate excited states $\ket{i}$, where $i = x,$ $y,$ $z$. The excited states are labeled by the direction of the dipole moment $\vect{d}_i = d \vect{e}_i$ associated with the transition $\ket{g} \to \ket{i}$, where $\vect{e}_i$ is the unit vector along the $i$ axis. We further define the atomic lowering operators $\hat{\sigma}_i = \ket{g}\bra{i}$ and the corresponding raising operators $\hat{\sigma}_i^\dagger = \ket{i} \bra{g}$. For an array of atoms, we require an additional index, $\hat{\sigma}_{mi}$, to label the site $m$ of the atom. By tracing out over the photonic degrees of freedom and making the Born--Markov approximation, one arrives at a quantum master equation describing the dynamics of the atoms~\cite{sLehmberg1970}. The master equation can be written in terms of the reduced density operator $\rho(t)$ as
\begin{equation}
  \frac{\mathrm{d}}{\mathrm{d} t} \rho(t) = - i \comm{H_S}{\rho(t)} + \mathcal{D}[\rho(t)],
\end{equation}
where
\begin{gather}
  H_S = \sum_{m, i} \omega_a \sigma_{m i}^\dagger \sigma_{m i} + \sum_{m \neq n,i,j} \Delta_{ij}(\vect{r}_m,\vect{r}_n) \sigma_{mi}^\dagger \sigma_{nj}
  \label{SM_eq:hamiltonian}
\end{gather}
captures the coherent dynamics, while
\begin{gather}
  \mathcal{D}[\rho] =  \sum_{m,n,i,j} \Gamma_{ij}(\vect{r}_m, \vect{r}_n) \left( \sigma_{mi}^\dagger \rho \sigma_{nj} - \frac{1}{2} \left\{ \sigma_{mi}^\dagger \sigma_{nj} , \rho \right\} \right)
\end{gather}
corresponds to incoherent, dissipative evolution.  The parameters $\Delta_{ij}(\vect{r}_m, \vect{r}_n)$ specify the coherent dipole--dipole interaction between two atoms, whereas $\Gamma_{ij}(\vect{r}_m, \vect{r}_n)$ denote the strength of their collective decay. These parameters can be expressed in terms of the dyadic Green's function of free space as
\begin{equation}
  \Delta_{ij}(\vect{r}_m, \vect{r}_n)= - \frac{3 \pi \gamma c}{\omega_a} \mathrm{Re} \left[ G_{ij} (\omega_a, \vect{r}_m, \vect{r}_n) \right], \qquad
  \Gamma_{ij}(\vect{r}_m, \vect{r}_n)= \frac{6 \pi \gamma c}{\omega_a} \mathrm{Im} \left[ G_{ij} (\omega_a, \vect{r}_m, \vect{r}_n) \right].
\end{equation}
We note the sum  involving $\Delta_{ij}(\vect{r}_m, \vect{r}_n)$ in Eq.~(\ref{SM_eq:hamiltonian}) excludes $m = n$ since this term merely corresponds to a renormalization of the transition frequency $\omega_a$~\cite{sdeVries1998,sKlugkist2006,sAntezza2009}.

For an infinite atomic array with discrete translational symmetry, it is convenient to introduce the momentum space operators
\begin{equation}
  \sigma_{\vect{k} i} = \sum_{m} \sigma_{m i} e^{i \vect{k} \cdot \vect{r}_m}, \qquad
  \sigma_{m i} = A_0 \int_{\mathrm{BZ}}\frac{\mathrm{d}^2 \vect{k}}{(2 \pi)^2} \sigma_{\vect{k} i} e^{- i \vect{k} \cdot \vect{r}_m}.
\end{equation}
where $\vect{k}$ is a 2D wavevector restricted to the first Brillouin zone of the reciprocal lattice and $A_0$ is the area of a unit cell. The Hamiltonian $H_S$ and the dissipator $\mathcal{D}$ may then be written as
\begin{gather}
  H_S = A_0 \int_{\mathrm{BZ}}\frac{\mathrm{d}^2 \vect{k}}{(2 \pi)^2} \sum_{i,j} \left[ \omega_a \delta_{ij} + \Delta_{ij}(\vect{k}) \right] \sigma_{\vect{k} i}^\dagger \sigma_{\vect{k} j} \label{SM_eq:hamiltonian_momentum}\\
  \mathcal{D}[\rho] =  A_0 \int_{\mathrm{BZ}} \frac{\mathrm{d}^2 \vect{k}}{(2 \pi)^2} \sum_{i,j} \left[ \gamma \delta_{ij} + \Gamma_{ij}(\vect{k}) \right] \left( \sigma_{\vect{k} i}^\dagger \rho \sigma_{\vect{k} j} - \frac{1}{2} \left\{ \sigma_{\vect{k} i}^\dagger \sigma_{\vect{k} j} , \rho \right\} \right).
\end{gather}
Here, we defined
\begin{equation}
  \Delta_{ij}(\vect{k}) = \sum_{m \neq n} \Delta_{ij}(\vect{r}_m, \vect{r}_n) e^{i \vect{k} \cdot (\vect{r}_m - \vect{r}_n)}, \qquad
  \Gamma_{ij}(\vect{k}) = \sum_{m \neq n} \Gamma_{ij}(\vect{r}_m, \vect{r}_n) e^{i \vect{k} \cdot (\vect{r}_m - \vect{r}_n)},
\end{equation}
independent of $n$ since $G(\omega_a, \vect{r}_m, \vect{r}_n)$ only depends on $\vect{r}_m - \vect{r}_n$. For a given momentum $\vect{k}$, the evolution of the atoms is thus specified by the two $3 \times 3$ matrices $\tensor{\Delta}(\vect{k})$ and $\tensor{\Gamma}(\vect{k})$. The former specifies the energies of the three Bloch modes with momentum $\vect{k}$, giving rise to a bandstructure, while the latter describes their decay.

The expressions for $\tensor{\Delta}$ and $\tensor{\Gamma}$ derived here are identical to those from the classical treatment. Of course, this is expected since the classical results should follow from the quantum treatment in the linear regime, where the atomic transitions are not saturated. Furthermore, the reflection and transmission coefficients computed from the linear response of the quantum system must agree with those obtained from the classical theory. One can show that this is indeed the case by introducing a driving term to the Hamiltonian, corresponding to an incident field, and computing the field radiated by the atoms.

\section{3. Properties of the cooperative resonance and decay}
\label{SM_sec_prop}
\subsection{3.1 Structure of the functions $\Delta(a)$ and $\Gamma(a)$ at normal incidence}
\label{SM_secKK}
Fig. 2a of the main text presents $\Delta$ as a function of the lattice spacing $a$ normalized to the incident wavelength $\lambda$.  The general features of its dependence on $a/\lambda$ can be elegantly explained by establishing its Kramers-Kronig relation with the function $\Gamma(a/\lambda)$. We recall the expressions for $\Delta$ and $\Gamma$ at normal incidence (c.f. Eq. 4 in the main text or Eq. \ref{SM_DG} here and text below it), which are in general functions of the incident frequency $\omega$ ($\lambda=2\pi c/\omega$) and the lattice points \begin{equation}
\Delta(\omega,a)=-\frac{3}{2}\gamma \lambda \sum_{n\neq 0} \mathrm{Re}G_{xx}(\omega,0,\mathbf{r}_n),
\quad
\Gamma(\omega,a)=3\gamma \lambda \sum_{n\neq 0} \mathrm{Im}G_{xx}(\omega,0,\mathbf{r}_n).
\label{SM_DGn}
\end{equation}
Since the dyadic Green's function is a linear response function (of the modes that form the electromagnetic field/vacum), it satisfies the Kramers-Kronig relation,
\begin{equation}
\mathrm{Re}G_{xx}(\omega,\mathbf{r},\mathbf{r}')=\frac{1}{\pi}\int_{-\infty}^{\infty}d\omega'\frac{\mathrm{Im}G_{xx}(\omega',\mathbf{r},\mathbf{r}')}{\omega-\omega'}.
\label{SM_KK}
\end{equation}
Writing the position vector of an atom $n$ as $\mathbf{r}_n=\mathbf{r}_{n_x,n_y}=a\mathbf{n}$ with $\mathbf{n}=(n_x\mathbf{e}_x+n_y\mathbf{e}_y)$ and considering the dydadic Green's function in free space, Eq. (\ref{SM_G}), we note that $\omega$ and $a$ always appear together as $2\pi\omega a/c=a/\lambda$ so that $\lambda G_{xx}(\omega,0,\mathbf{r}_n)$ can be written as a function of only the dimensionless distance $a/\lambda$ and the vector index $\mathbf{n}$ (independent of $a$ and $\omega$). Therefore, the Kramers-Kronig relation, Eq. (\ref{SM_KK}), for $G_{xx}(\omega,0,\mathbf{r}_n)$ can be written as
\begin{equation}
\mathrm{Re}[\lambda G_{xx}(a/\lambda,\mathbf{n})]=\frac{1}{\pi}\int_{-\infty}^{\infty}du\frac{\mathrm{Im}[\lambda G_{xx}(u,\mathbf{n})]}{a/\lambda-u}.
\label{SM_KKu}
\end{equation}
Inserting this relation into the expression for $\Delta$ in Eq. (\ref{SM_DGn}) and performing the sum over $n$, the dependence on $\mathbf{n}$ disappears and we obtain a Kramers-Kronig relation between the real and imaginary part of the field response as a function of the scaled lattice spacing $a/\lambda$
\begin{equation}
\Delta(a/\lambda)=-\frac{1}{\pi}\int_{-\infty}^{\infty}du\frac{\Gamma(u)/2}{a/\lambda-u},
\label{SM_KKDG}
\end{equation}
where both $\Delta(u)$ and $\Gamma(u)$ are now understood to be functions of a single variable $u=a/\lambda$.
\begin{figure}[t]
  \centering
  \includegraphics[scale=0.4]{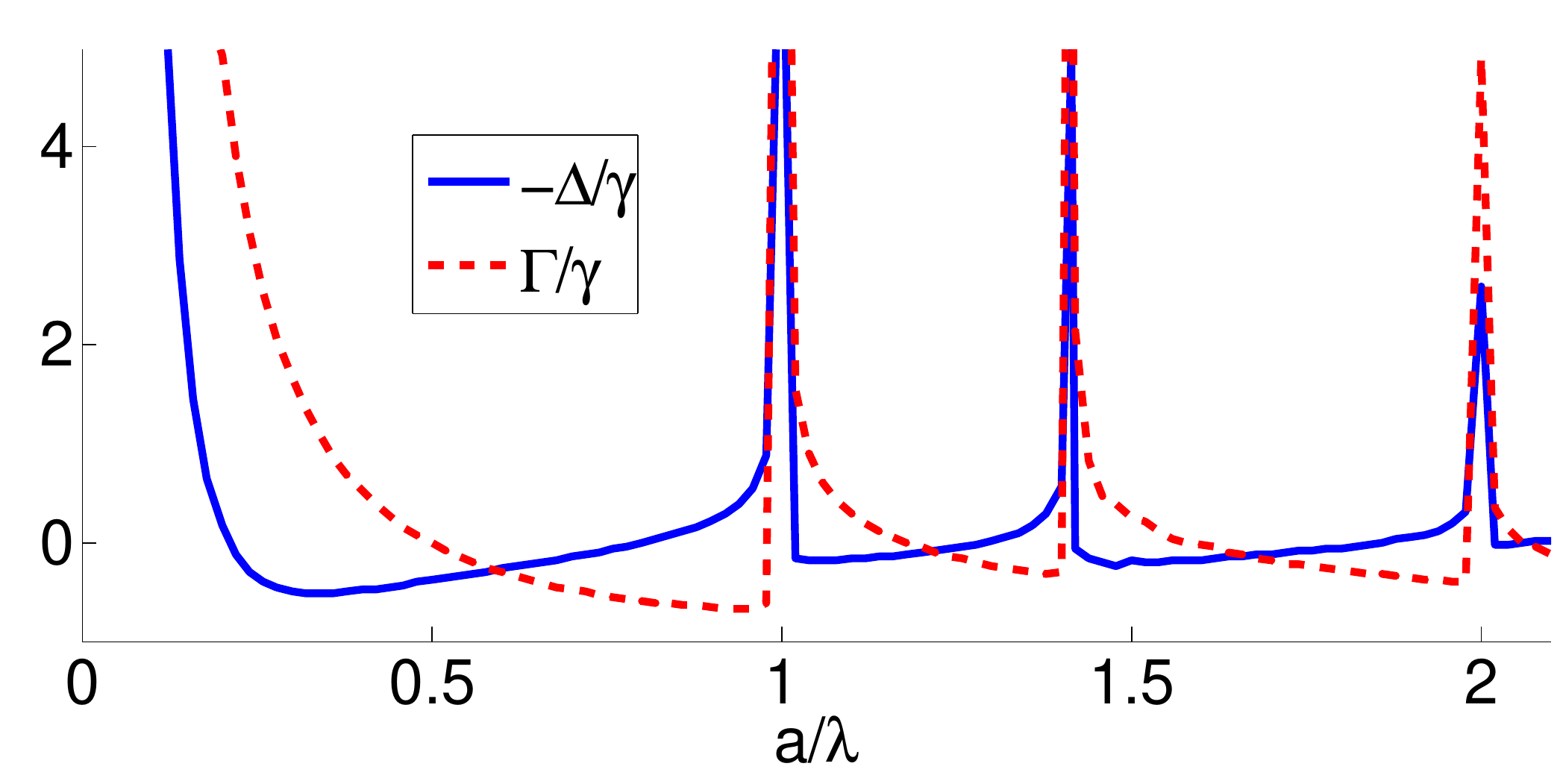}
\caption{\small{
Cooperative shift $\Delta$ and width $\Gamma$ as a function of $a/\lambda$ at normal incidence (extension of Fig. 2a in the main text; results obtained by a direct summation of Eq. (4) in the main text). The diffraction-orders-associated peaks and the Kramers-Kronig structure of the relation between the $\Delta$ and $\Gamma$ as a function of the scaled lattice constant $a/lambda$ are clearly seen (see text, Sec. \ref{SM_secKK}).
 }} \label{SM_figS1}
\end{figure}

Turning now to the structure of the function $\Delta(a/\lambda)$ we begin with the region $a/\lambda\ll 1$. In this limit the sum over $n$ in Eq. (\ref{SM_DGn}) can be converted into an integral and we analytically find $\Delta\propto 1/a^3$ which agrees with the divergence near $a/\lambda \rightarrow 0^+$ in Fig. 2a of the main text and Fig. S1 here (see also Ref. \cite{sGDA2}). This $1/a^3$ scaling is easily understood by recalling that the real part of the Green's function Eq. (\ref{SM_G}), amounting to dipole-dipole interaction, scales as $1/r^3$ at the quasistatic, short distance limit, and that $\Delta$ is a sum over such dipole-dipole interactions. In order to explain the behavior of  $\Delta(a/\lambda)$ at all regions beyond the $a/\lambda\ll 1$ we reside to the Kramers-Kronig relation Eq. (\ref{SM_KKDG}), for which we need to consider the function $\Gamma(a/\lambda)$. For $a/\lambda<1$ we have obtained an analytical expression for $\Gamma$ (Eq. 4 in main text) by using the fact that there exists only one diffraction order in the scattered light. In other words, the reservoir established by the free-space vacuum, forms only a single radiative \emph{dissipation channel} for the atomic array at $a/\lambda<1$, imposing a dissipation $\Gamma$ which results only from the term $m_x,m_y=0$ in Eq. (\ref{SM_g3}). For $a/\lambda=1$, additional dissipation channels arise in the form of emission to the directions (diffraction orders) $|m_x|=1, m_y=0$ and $|m_y|=1, m_x=0$. This physically explains the peak observed in $\Gamma(a/\lambda)$ for $a/\lambda=1$ in Fig. S1 (mathematically, additional poles contribute to the imaginary part of the sum in Eq. \ref{SM_g3}). The meaning is that whenever a new dissipation channel appears, in the form of a new diffraction order in the far field, we expect a peak in  $\Gamma(a/\lambda)$. Indeed, this can be seen in Fig. S1 by the additional peaks at $a/\lambda=\sqrt{2}$ (diffraction orders $|m_x|=|m_y|=1$) and $a/\lambda=2$ (diffraction orders $|m_x|=2,m_y=0$ and $|m_y|=2,m_x=0$). In turn, these peaks in $\Gamma(a/\lambda)$ then physically explain the corresponding dispersive-like peaks in $\Delta(a/\lambda)$ following the Kramers-Kronig relation Eq. (\ref{SM_KKDG}), nicely seen in Fig. S1 (whereas Fig. 2a in the main text captures only the $a/\lambda=1$ "dispersive" peak).

\subsection{3.2 Polarization eigenbasis of $\tensor{\Delta}$ and $\tensor{\Gamma}$}
For $a < \lambda/2$, we derived an explicit expression for $\tensor{\Gamma}(\vect{k}_\parallel)$ in Eq.~(\ref{SM_Ga}). Written in terms of spherical coordinates $\theta$ and $\phi$, the expression becomes
\begin{equation}
  \tensor{\Gamma}(\vect{k}_\parallel) =
  \frac{3 \gamma}{4 \pi \cos \theta} \left( \frac{\lambda}{a} \right)^2
  \begin{pmatrix}
    1 - \sin^2 \theta \cos^2 \phi & - \sin^2 \theta \cos \phi \sin \phi & 0\\
    - \sin^2 \theta \cos \phi \sin \phi & 1 - \sin^2 \theta \sin^2 \phi & 0\\
    0 & 0 & \sin^2 \theta
  \end{pmatrix}
  - \gamma
  \begin{pmatrix}
    1 & 0 & 0 \\
    0 & 1 & 0 \\
    0 & 0 & 1
  \end{pmatrix} .
\end{equation}
The matrix can be straightforwardly diagonalized. One obtains the three eigenvalues
\begin{equation}
  \Gamma_1 = \frac{3 \gamma \cos \theta}{4 \pi} \left( \frac{\lambda}{a} \right)^2 - \gamma, \qquad
  \Gamma_2 = \frac{3 \gamma}{4 \pi \cos \theta} \left( \frac{\lambda}{a} \right)^2 - \gamma, \qquad
  \Gamma_3 = \frac{3 \gamma \sin^2 \theta}{4 \pi \cos \theta} \left( \frac{\lambda}{a} \right)^2 - \gamma, \qquad
\end{equation}
with the respective eigenvectors
\begin{equation}
  \vect{v}_1 = ( \cos \phi , \sin \phi , 0 ) = \vect{k}_\parallel / |\vect{k}_\parallel|, \qquad
  \vect{v}_2 = ( -\sin \phi ,  \cos \phi , 0 ) = \vect{e}_s, \qquad
  \vect{v}_3 = ( 0 , 0, 1 ) = \vect{e}_z .
\end{equation}
We observe that $\vect{v}_2$ points along the direction of $s$-polarized incident light, while $\vect{v}_1$ and $\vect{v}_3$ are superpositions of the $p$-polarization and a longitudinal component.

\begin{figure}[h]
  \centering
  \includegraphics[scale=0.6]{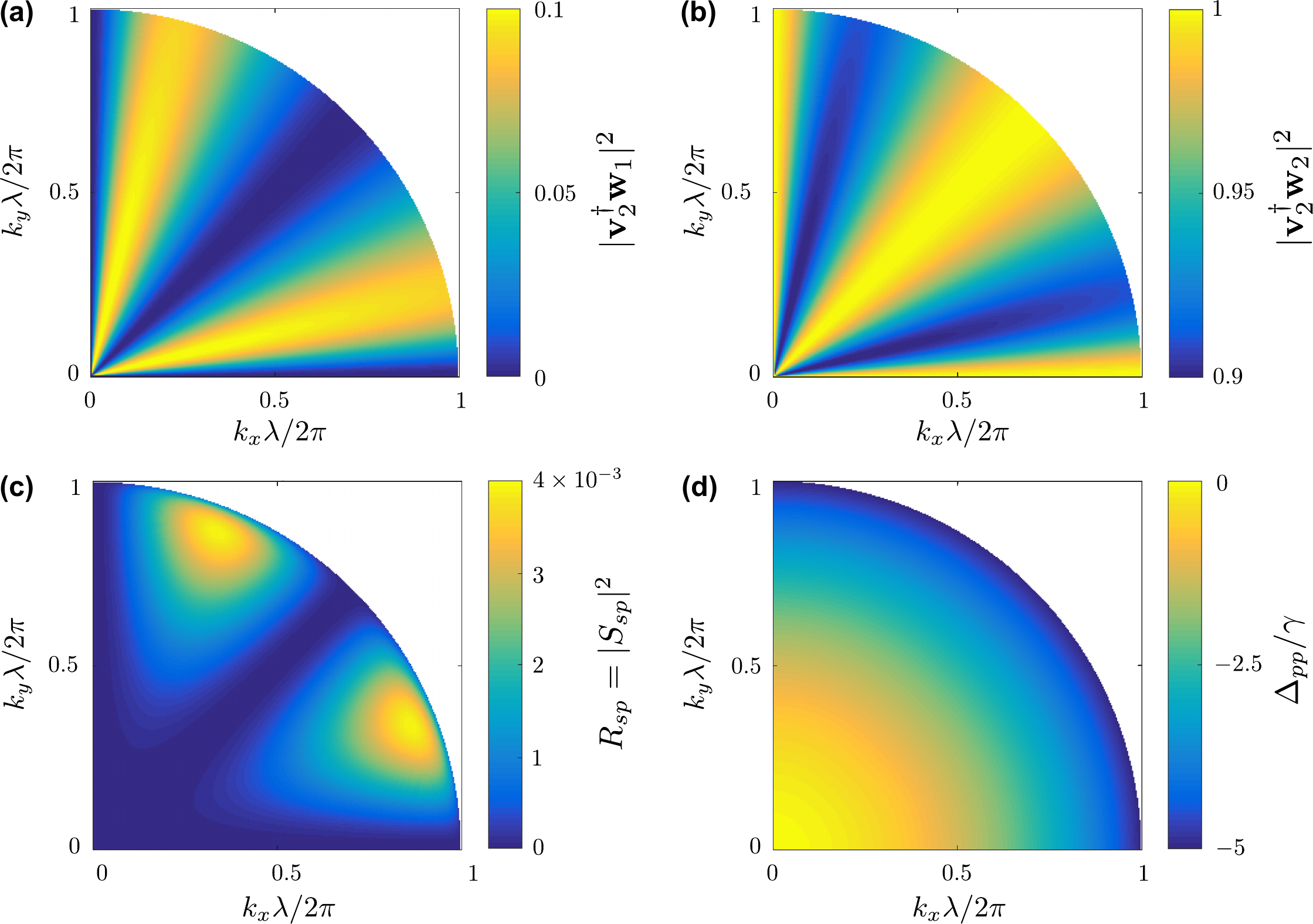}
  \caption{Overlap between the two in-plane eigenvectors $\vect{w}_1$ and $\vect{w}_2$ of $\tensor{\Delta}$ and one of the in-plane eigenvectors $\vect{v}_2$ of $\tensor{\Gamma}$. (a) indicates that $\vect{w}_1$ is approximately orthogonal to $\vect{v}_2$ for all $\vect{k}_\parallel$ inside the light cone. Similarly, it is evident from (b) that $\vect{w}_2$ is approximately parallel to $\vect{v}_2$. (c) Off-diagonal reflection coefficient $R_{sp}$ (due to symmetry, $R_{sp} = R_{ps}$). (d) Matrix element $\Delta_{pp}(\vect{k}_\parallel) = \vect{e}^{- \dagger}_p \tensor{\Delta}(\vect{k}_\parallel) \vect{e}^{+}_p$ of the cooperative shift $\tensor{\Delta}(\vect{k}_\parallel)$. All plots were computed for lattice constant $a/\lambda = 0.2$.}
  \label{SM_fig:overlap}
\end{figure}

The cooperative shift $\tensor{\Delta}(\vect{k}_\parallel)$ has the same block-diagonal structure as $\tensor{\Gamma}(\vect{k}_\parallel)$, such that $\vect{v}_3 = ( 0, 0, 1)$ is also an eigenvector of $\tensor{\Delta}(\vect{k}_\parallel)$. The other two eigenvectors, which we denote by $\vect{w}_{1,2}$, are in general different. However, it turns out that they are approximately equal to $\vect{v}_{1,2}$ as demonstrated in Fig.~\ref{SM_fig:overlap}. As a result, the $s$-polarization is an approximate eigenvector of both $\tensor{\Gamma}$ and $\tensor{\Delta}$ and hence of the scattering matrix. This further implies that the surface only weakly mixes the $s$- and $p$-polarizations. Indeed, for $a / \lambda = 0.2$, the reflection coefficient $R_{sp}$ never exceeds $4 \times 10^{-3}$ as shown in Fig.~\ref{SM_fig:overlap}c ($R_{sp} = R_{ps}$ by symmetry).

The above discussion of the eigenvectors of $\tensor{\Delta}$ and $\tensor{\Gamma}$ also explains why the matrix element $\Delta_{ss}$ alone is sufficient to predict the reflection coefficient $R_{ss}$ to a good approximation (see main text). By contrast, no such simple explanation is available for the $p$-polarization since it forms a superposition of $\vect{w}_1$ and $\mathbf{e}_z$.
Nevertheless, for small enough incident angles within the paraxial regime where  $\mathbf{e}^{\pm}_p\approx\pm\mathbf{k}_{\parallel}/|\mathbf{k}_{\parallel}|\approx \pm\vect{w}_1$, the $p$-polarization becomes an approximate eigenvector of $\tensor{\Gamma}$ and $\tensor \Delta$. The plot of $\Delta_{pp} = \vect{e}^{- \dagger}_p \tensor{\Delta} \vect{e}^{+}_p$  in Fig.~\ref{SM_fig:overlap}d then exhibits a resonance ($\Delta_{pp}\approx0=\delta$) matching that of $R_{pp}$ within the paraxial regime, but fails to reproduce many qualitative features of $R_{pp}$ beyond it (see Fig. 3c of the main text).

\section{4. Disordered arrays}
\label{SM_sec_dis}
Disorder in atomic positions can be described by writing $\mathbf{r}=\mathbf{r}^0_n+\delta\mathbf{r}_n$ for the position of an atom $n$, where $\mathbf{r}^0_n$ are the ordered lattice positions and $\delta\mathbf{r}_n$ are a small random fluctuations. The effect on the self-consistent scattering equation, Eq. (\ref{SM_LSp}), is a perturbation $\tensor{\delta G}_{nm}$ of the Green's function,
\begin{equation}
\tensor{G}_{nm}=\tensor{G}(\mathbf{r}_n,\mathbf{r}_m)=\tensor{G}(\mathbf{r}^0_n,\mathbf{r}^0_m)+\tensor{\delta G}_{nm}, \quad \tensor{\delta G}_{nm}\approx \sum_{a=n,m}\sum_{i=x,y,z}\frac{\partial \tensor{G}(\mathbf{r}_n,\mathbf{r}_m)}{\partial r_a^i}\delta r_a^i+...
\label{SM_delG}
\end{equation}
with $\delta r_n^i=\mathbf{e}_i\cdot \delta\mathbf{r}_n^i$. We then derive a formal perturbation theory for Eq. (\ref{SM_LSp}), finding that to lowest order the effect on the effective polarizability is a correction to the cooperative resonance,
\begin{equation}
\tensor{\delta\Delta}(\mathbf{k}_{\parallel})=-\frac{3}{2}\gamma \lambda \frac{1}{N}\sum_n\sum_{m\neq n} \mathrm{Re}\left[\overline{\overline{\delta G}}_{nm}e^{-i\mathbf{k}_{\parallel}\cdot(\mathbf{r}_n-\mathbf{r}_m)}\right].
\end{equation}
For a simple estimation of the effect of disorder, we take statistically independent and identically-distributed fluctuations $\delta r_n^i$ with zero mean and variance $\delta r^2$,
\begin{equation}
\langle \delta r_n^i\rangle =0, \quad \langle \delta r_n^i \delta r_m^j \rangle=\delta r^2 \delta_{nm} \delta_{ij}.
\end{equation}
Considering the Taylor expansion of $\overline{\overline{\delta G}}_{nm}$ from Eq. (\ref{SM_delG}) up to second order (valid for $\delta r\ll a$), we find e.g., that at normal incidence, the average of the disorder-induced shift is given by $\langle \delta\Delta \rangle\approx  4\pi^2 (\delta r/\lambda)^2 \Delta$.

For disorder due to impurities, we consider the case of an absence or addition of an atom in the array. We find numerically (see Sec. \ref{SM_sec_num} below) that for small values of $a/\lambda$ (e.g. $0.2$) the results are almost unchanged compared with those of a perfect lattice, whereas for $a/\lambda$ larger than $1/2$ (e.g. $0.8$) they may change considerably. This may support the requirement for a high filling factor found in \cite{sADM1} for the case $a/\lambda=0.8$ ($\delta=0$).

\section{5. Optical saturation of the array}
\label{SM_sec_NLO}
Here we address the optical nonlinearity of the array by estimating the incident power needed to saturate individual atoms. We first evaluate he fraction of incident power that is absorbed (and then scattered) in each individual atom of the array during the scattering process at the collective resonance ($\delta=\Delta$). This allows us to asses the number of atoms that actually participate in the interaction and scattering of the incident field, hinting at the potential degree of nonlinearity of the array (main text).

We consider a normal incident Gaussian beam,
\begin{equation}
\mathbf{E}_0(\mathbf{r})=E_0\mathbf{e}_L\frac{w_0}{w(z)}e^{ikz}e^{-i\varphi(z)}e^{-\frac{x^2+y'^2}{w^2(z)}}e^{ik\frac{x^2+y^2}{2R(z)}},
\label{SM_Gau1}
\end{equation}
with the usual beam parameters,
\begin{equation}
w(z)=w_0\sqrt{1+\left(\frac{z}{z_R}\right)^2}, \quad z_R=\frac{\pi w_0^2}{\lambda}, \quad R(z)=z\left[1+\left(\frac{z_R}{z}\right)^2\right], \quad \varphi(z)=\arctan\left(\frac{z'}{z_R}\right).
\label{SM_Gau2}
\end{equation}
and where $w_0$ is the beam waist at its focal point and $\mathbf{e}_L$ its polarization.
The power carried by the beam is
\begin{equation}
W=\frac{\pi}{2}w_0^2 c\epsilon_0|E_0|^2.
\label{SM_W}
\end{equation}
 Considering the paraxial character of the incident Gaussian beam, namely, that $|\mathbf{k}_{\parallel}|/k\ll 1$ for all $\mathbf{k}_{\parallel}$ contained in $\mathbf{E}_{0,\mathbf{k}_{\parallel}}$, we have $\tensor{\Gamma}(\mathbf{k}_{\parallel})\approx \tensor{\Gamma}(0)$ and $\tensor{\Delta}(\mathbf{k}_{\parallel})\approx \tensor{\Delta}(0)$ [see e.g. Eq. (\ref{SM_Ga}) and Fig. S2d here or Fig. 3c of the main  the main text], so that $\tensor{\alpha}_e(\mathbf{k}_{\parallel})\approx\tensor{\alpha}_e(0)=\alpha_e\delta_{ij}$
and Eq. (\ref{SM_p}) becomes
\begin{equation}
\mathbf{p}(\mathbf{k}_{\parallel})\approx \alpha_e \mathbf{E}_{0,\mathbf{k}_{\parallel}} \quad \Rightarrow \quad \mathbf{p}_n\approx \alpha_e \mathbf{E}_{0}(\mathbf{r}_n)
\label{SM_pp}
\end{equation}
Therefore, within the paraxial approximation around normal incidence, it appears as if each atom individually possess an effective polarizability $\alpha_e$, such that the power dissipated in an atom $n$ (solely by radiation/scattering for the case $\gamma_{\mathrm{nr}}=0$) at cooperative resonance, $\delta=\Delta$, is given by
\begin{equation}
W_n=W_{n_x,n_y}=\frac{1}{2}\mathrm{Im}[\alpha_e] \omega|\mathbf{E}_0(\mathbf{r}_n)|^2=\epsilon_0 c|E_0|^2a^2 e^{-2\frac{a^2}{w_0^2}(n_x^2+n_y^2)},
\label{SM_Wn}
\end{equation}
where $\Gamma+\gamma=\gamma[3/(4\pi)](\lambda/a)^2$ was used in the expression for $\alpha_e$ at normal incidence (Eq. \ref{SM_alpen}). It can be verified that at the cooperative resonance all of the incident power is absorbed (scattered) by the array, by summing over all atoms $(n_x,n_y)$ and arriving at $\sum_{n_x}\sum_{n_y}W_{n_x,n_y}\approx \int_{-\infty}^{\infty} dn_x \int_{-\infty}^{\infty} dn_y W_{n_x,n_y}=W$ with $W$ from Eq. (\ref{SM_W}).

The fraction of power absorbed by an individual atom is then given by
\begin{equation}
P_n=\frac{W_n}{W}=\frac{a^2}{(\pi/2)w_0^2} e^{-2\frac{a^2}{w_0^2}(n_x^2+n_y^2)}.
\label{SM_Pn}
\end{equation}
Therefore, the atoms that are within the beam waist, namely, those which significantly participate in the interaction with light and its scattering, absorb a fraction of $P_n=a^2/[(\pi/2)w_0^2]$ of the incoming power (per atom). By noticing that the cross section (area) of the Gaussian beam scales as $\sim(\pi/2)w_0^2$, the result for $P_n$ can be interpreted as a ratio between atomic and light cross sections, with the atomic effective cross section given by $\sim a^2$. This interpretation is also supported by comparing the polarizabilities  at resonance of bare and dressed atoms: from Eq. (\ref{SM_alp}) and Eq. (\ref{SM_alpen}) we arrive at $\alpha(\delta=0)\propto\lambda^3$ and  $\alpha_e(\delta=\Delta)\propto \lambda a^2$  which demonstrates how the individual cross section $\lambda^2$ of a bare atom is replaced by the effective cross section $a^2$ for the renormalized/dressed atom.

Since atoms are highly nonlinear, namely, they become transparent to light upon their full excitation (saturation), then the mirror effect may vanish for sufficiently high incident power $W$. In order to estimate the power required for the saturation of the atoms, we note that an atom $n$ is likely to be excited upon absorbing $N=1/P_n$ photons at a time $<(\Gamma+\gamma)^{-1}$ before it decays, setting a saturation power of  $ W\approx N\hbar\omega(\Gamma+\gamma)$. Taking, e.g., $a=0.49\lambda$ and $w_0=1.5\lambda$ we obtain $\Gamma+\gamma\approx\gamma$ and $N\approx 14$ for atoms within the beam waist, so that $W\approx 14\hbar \omega \gamma$, only a single order of magnitude larger than the saturation power of a single atom coupled to a 1d photonic channel.

\section{6. Direct numerical approach to the scattering problem}
\label{SM_sec_num}
The problem of scattering of an electromagnetic field by a finite collection of point dipoles, generally formulated by Eqs. (\ref{SM_LS}) and (\ref{SM_LSp}) can be solved numerically by a simple matrix inversion. Introducing the vector notation $\overline{E}$ for the $3N$-dimensional vector $E^n_i=p^n_i/\alpha$ (3 first entries are the 3 vector components of the $E$-field at the position of atom 1, next 3 are the field components at atom 2, etc.) and the notation $\overline{\overline{\mathbb{G}}}$ for the $3N\times3N$ matrix $G^{nm}_{ij}$, the solution of Eq. (\ref{SM_LSp}) for the local fields on the atoms, $\overline{E}=E_i(\mathbf{r}_n)$, is
\begin{equation}
\overline{E}=\left[1-4\pi^2\frac{\alpha}{\epsilon_0\lambda^3}\lambda \overline{\overline{\mathbb{G}}}\right]^{-1}\overline{E}_0.
\label{SM_sol1}
\end{equation}
Given the collection of atomic positions that form the array $\mathbf{r}_n$ (e.g. square lattice), the incident field $\mathbf{E}_0(\mathbf{r})$ (see below) and the polarizability $\alpha(\omega)$ [Eq. \ref{SM_alp}], we can numerically invert the matrix in Eq. (\ref{SM_sol1}).
Then, we insert the solutions $\overline{E}=E^n_i=E_i(\mathbf{r}_n)$ into the right-hand side of Eq. (\ref{SM_LS}) and obtain the solution for the field at any given point $\mathbf{r}$, $E_i(\mathbf{r})$.

In order to compare the results of our analytical theory to those obtained numerically in a more realistic scenario, the infinite array is replaced by a finite square lattice of $N$ atoms whereas all incident plane waves are replaced by Gaussian beams,
\begin{equation}
\mathbf{E}_0(x',y',z')=E_0\mathbf{e}_L\frac{w_0}{w(z')}e^{ikz'}e^{-i\varphi(z')}e^{-\frac{x'^2+y'^2}{w^2(z')}}e^{ik\frac{x'^2+y^2}{2R(z')}},  \label{SM_Gau1a}
\end{equation}
with the beam parameters from Eq. (\ref{SM_Gau2}).
Here the coordinates $(x',y',z')$ are written in the reference frame of the beam, namely, where the beam propagates along the $z'$ direction, $\mathbf{e}_k=\mathbf{e}_{z'}$. For normal incident Gaussian beam we have $x=x',y=y',z=z'$ and a polarization $\mathbf{e}_L\in\{\mathbf{e}_{x},\mathbf{e}_{y}\}$, whereas for a beam propagating along the $xz$-plane at an angle $\theta$ from the $z$ axis, we find
\begin{equation}
x'=x\cos\theta-z\sin\theta, \quad y'=y, \quad z'=z\cos\theta+z\sin\theta,
\label{SM_Gau3}
\end{equation}
and
\begin{equation}
\mathbf{e}_L\in\{\mathbf{e}_{p},\mathbf{e}_{s}\},\quad \mathrm{with} \quad \mathbf{e}_p=\cos\theta\mathbf{e}_x-\sin\theta\mathbf{e}_z, \quad \mathbf{e}_s=-\mathbf{e}_y.
\label{SM_Gau4}
\end{equation}
For a faithful comparison with the infinite array case assumed in the theory, the cross section of the Gaussian beam at the position of the array, $z=0$, has to be sufficiently smaller than the area of the array, $Na^2$. At normal incidence ($\theta=0$) we then always take $w_0\leq 0.3 a\sqrt{N}$, whereas for a general angle $\theta$ the waist $w_0$ has to be smaller by roughly $\cos\theta$.

Figs. S3 and S4 illustrate the excellent qualitative and quantitative agreement between this numerical solution and the theoretical results presented in the main text, both for the normal incident and oblique incident cases, respectively.

\begin{figure}
\begin{center}
\includegraphics[scale=0.4]{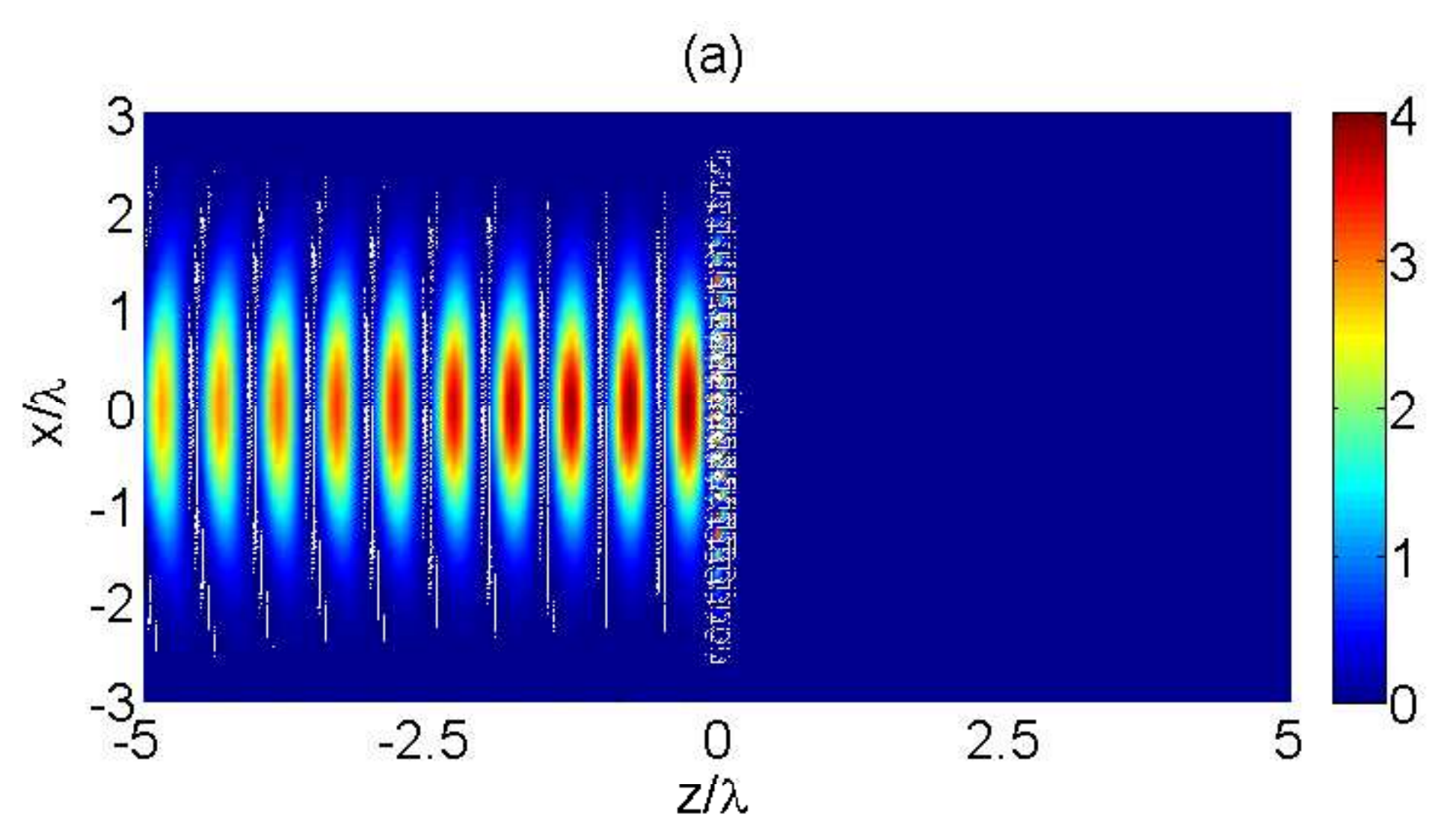}
\includegraphics[scale=0.4]{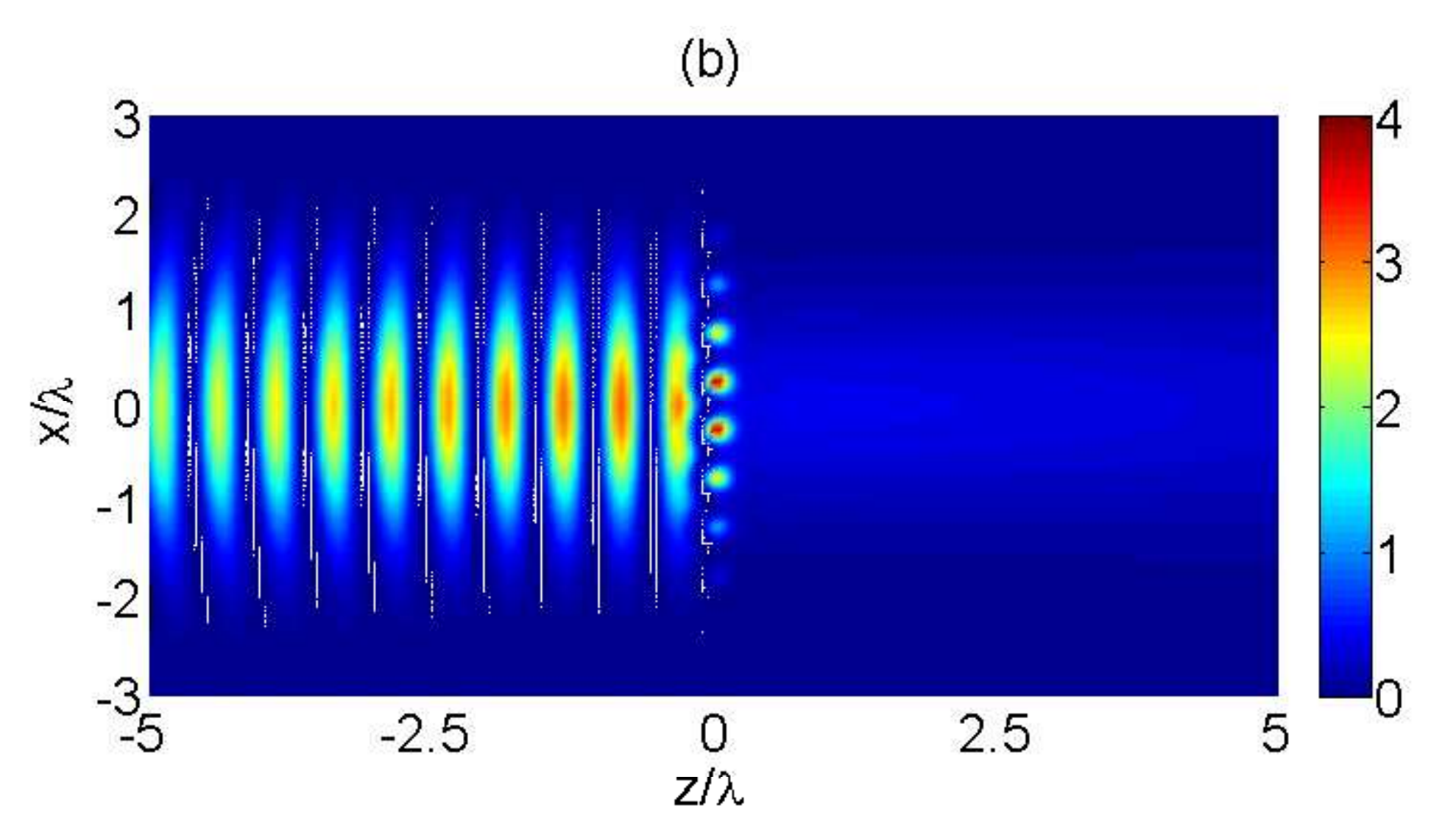}
\includegraphics[scale=0.4]{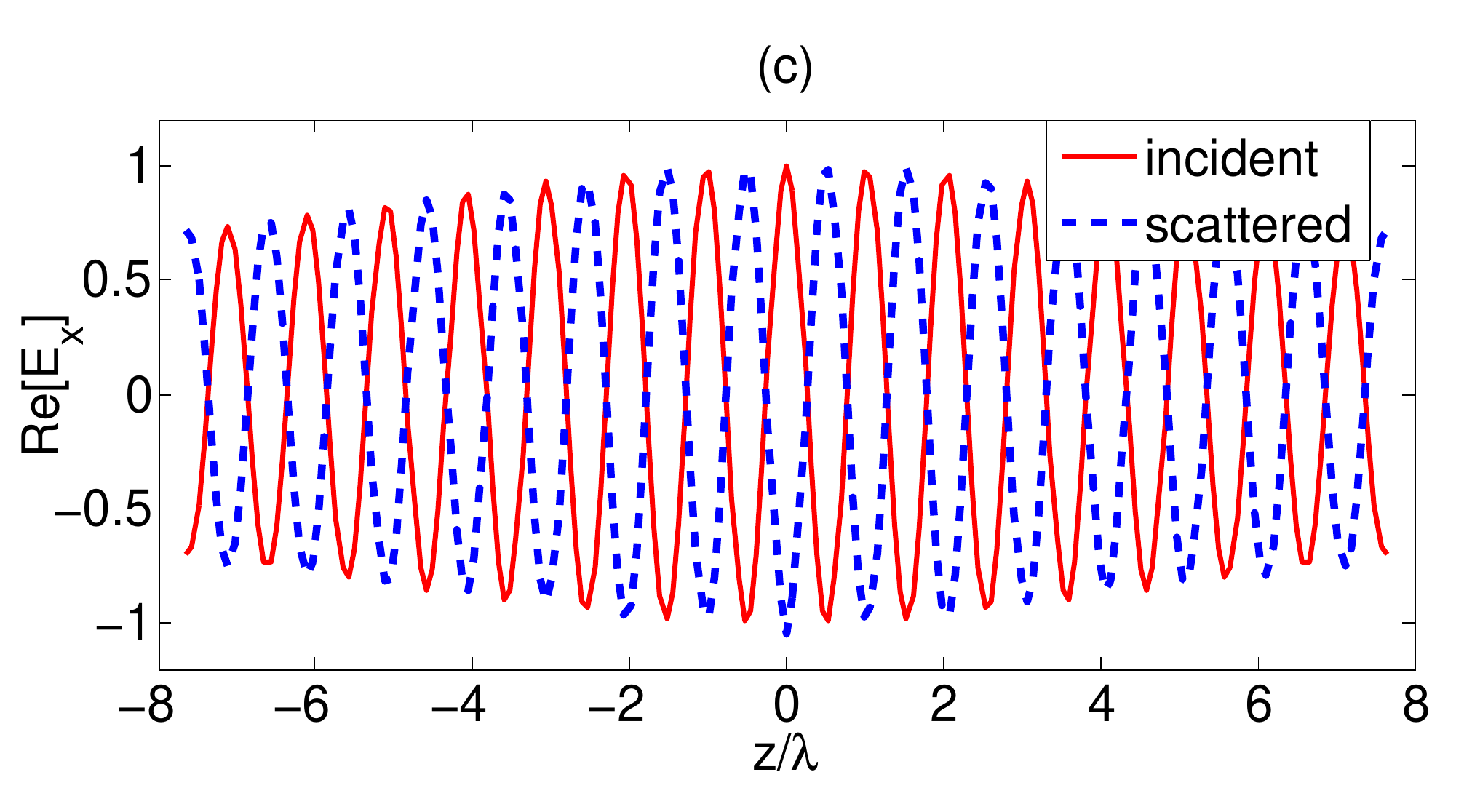}
\includegraphics[scale=0.4]{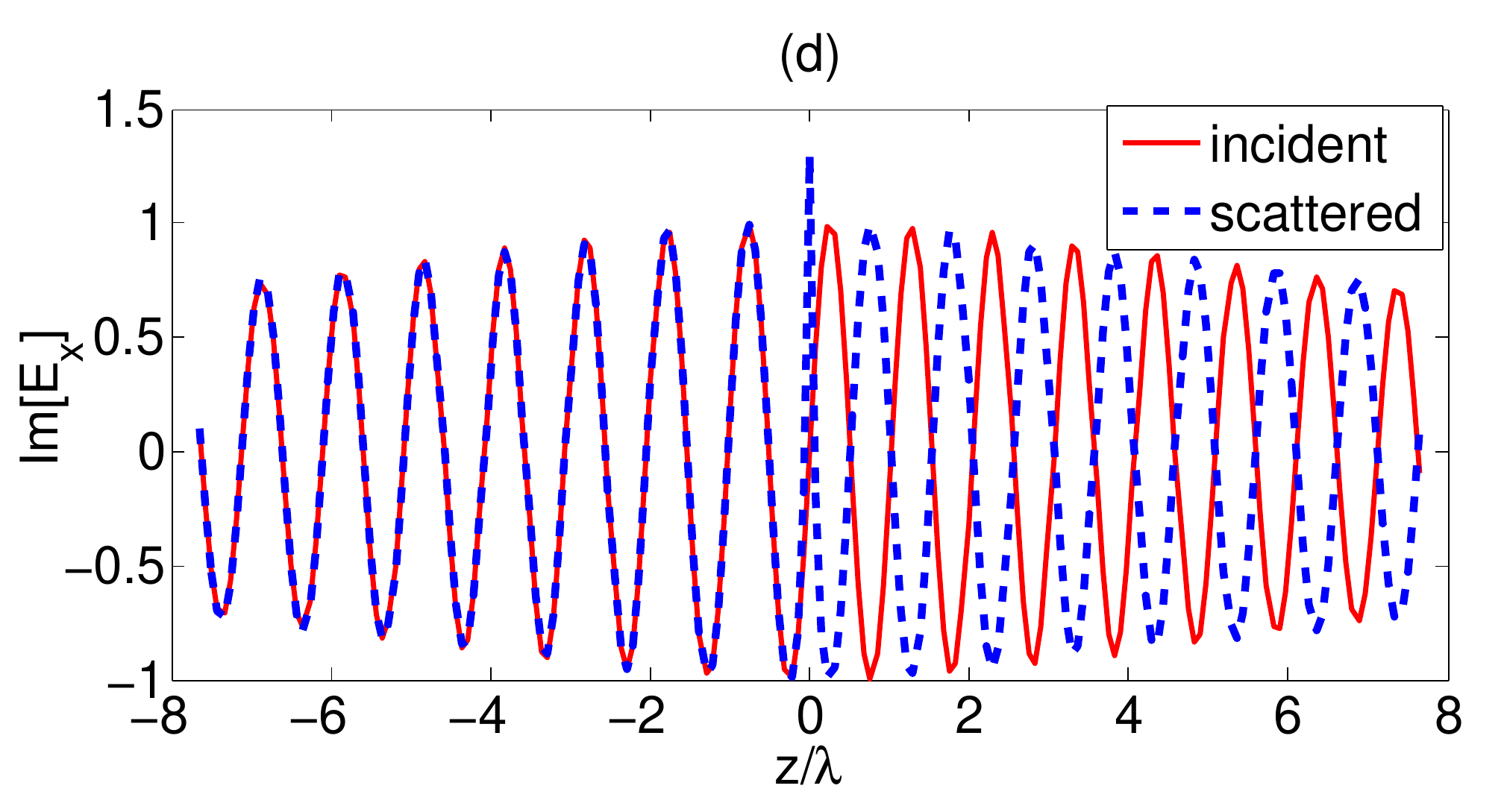}
\includegraphics[scale=0.4]{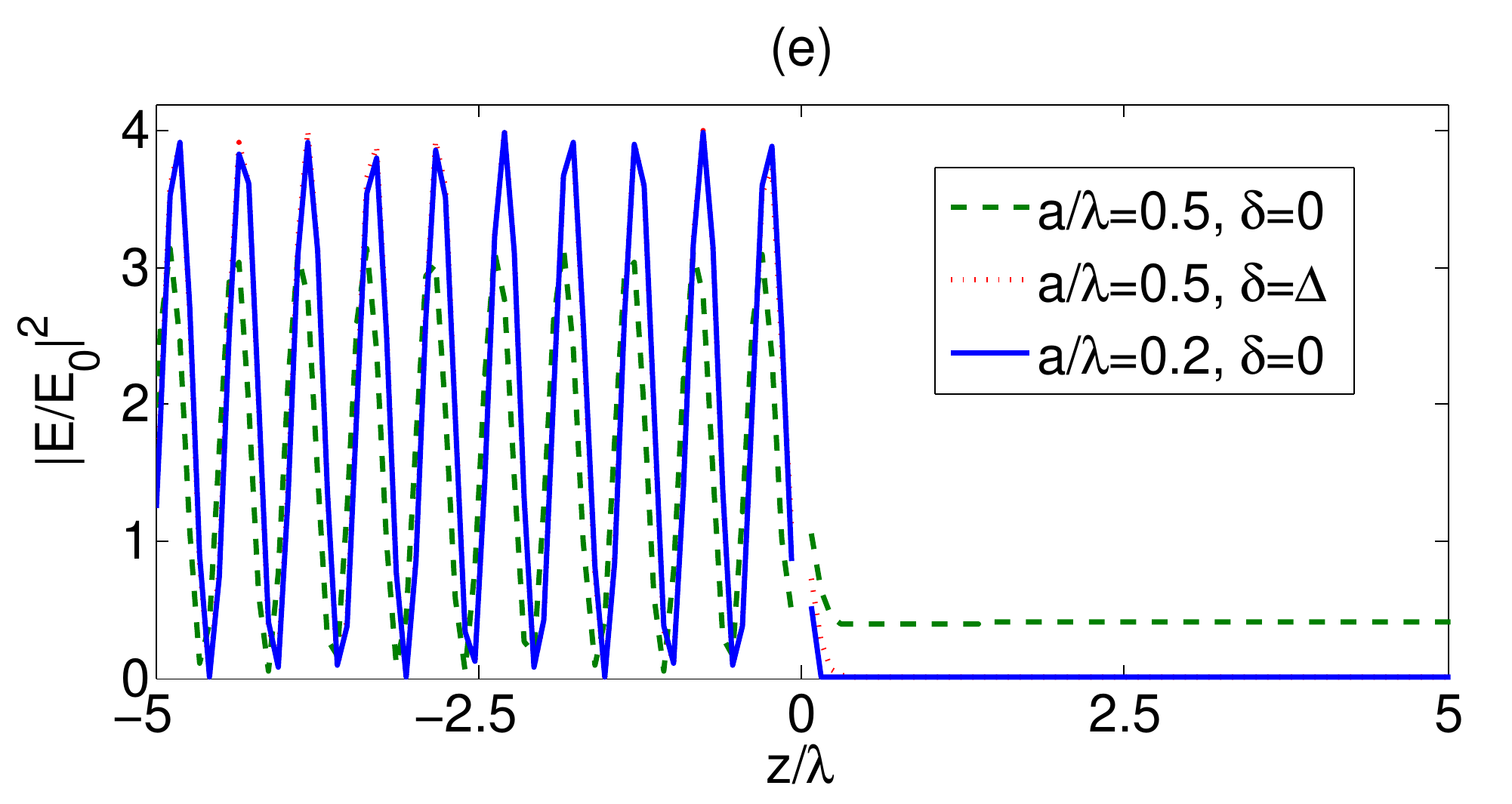}
\includegraphics[scale=0.4]{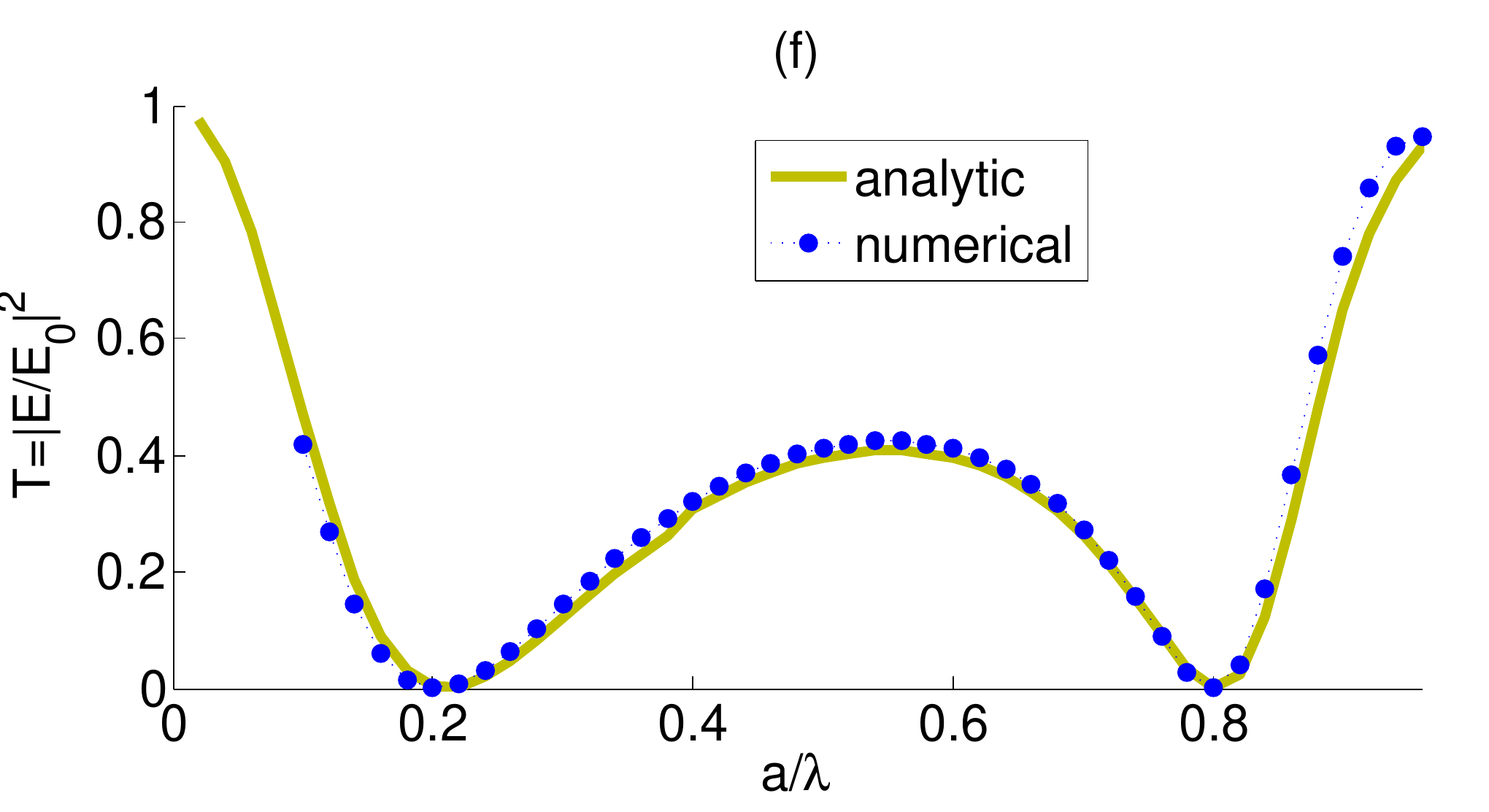}
\caption{\small{
Numerical approach, normal incidence. (a) $x$-polarized incident Gaussian beam at normal incidence to an array of $N=26\times26$ atoms with a lattice constant $a=0.2\lambda$ at $z=0$. The incident beam is resonant with the bare atom, $\delta=0$, and its waist is $w_0=0.3\sqrt{N} a=1.56\lambda$. In agreement with Fig. 1c of the main text, no transmission is observed, and the reflected wave forms a standing wave with the incident field. All fields intensities are in units of $|E_0|^2$, the peak intensity of the incident Gaussian beam, Eq. (\ref{SM_Gau1a}). (b) Same as (a) for $a=0.5$ and $w_0=0.3\sqrt{N} \times 0.2=1.56\lambda$. Here it appears that some of the field is transmitted since as in Fig. 1c (main text). (c) A closer look at the scattering and interference processes for the case plotted in (a). Here we plot the incident and scattered field along $z$ for a constant arbitrary value of $x$ and $y$ ($x=y=0.2a$). The real part of the scattered field is exactly opposite in sign with respect to the real part of the incident field thus exactly cancelling it (all fields in units of $E_0$). (d) Same as (c) for the imaginary part of the field. For $z>0$, beginning at a short distance after evanescent fields have decayed, there exists again exact cancellation, so that no transmitted (real and imaginary) field exists. For $z<0$ the scattered field is equal and in phase with the incident field, amounting for perfect reflection and a standing wave. (e) Same as (c), this time the blue solid curve presents the total field normalized to the incident one, showing the standing wave due to perfect reflection and zero transmission. The dashed green curve presents the same calculation for $a=0.5\lambda$ where at $\delta=0$ a transmission of about $\sim 0.4$ is observed in agreement with Fig. 1c of the main text. However, vanishing transmission (perfect reflection) can be achieved also at $a=0.5\lambda$ by changing the frequency of the incident field to match the cooperative resonance $\delta=\Delta$, which is found for $a=0.5\lambda$ using Fig. 2a in the  main text. The result is presented by the red dotted line which almost exactly coincides with the blue curve apart from the somewhat longer evanescent tail at $z>0$ (which grows with $a$). (f) Transmitted field as a function of lattice spacing $a/\lambda$. Repeating the calculation from (d) for different $a$ values at the individual-atomic resonance, $\delta=0$, the transmission coefficient is extracted. For each $a$ value the Gaussian beam waist is taken to be $w_0=0.3\sqrt{n}a$, i.e. smaller than the size of the array (blue dots). This nevertheless limits $a/\lambda$ to be above $0.1$ in order to keep $w_0>\lambda/2$ (for the paraxial approximation to be valid). Excellent agreement with the analytical calculations for an infinite array and a plane wave, Fig. 1c in the main text which is reproduced here (green solid curve), is observed.
 }} \label{SM_fig3}
\end{center}
\end{figure}

\begin{figure}
\begin{center}
\includegraphics[scale=0.4]{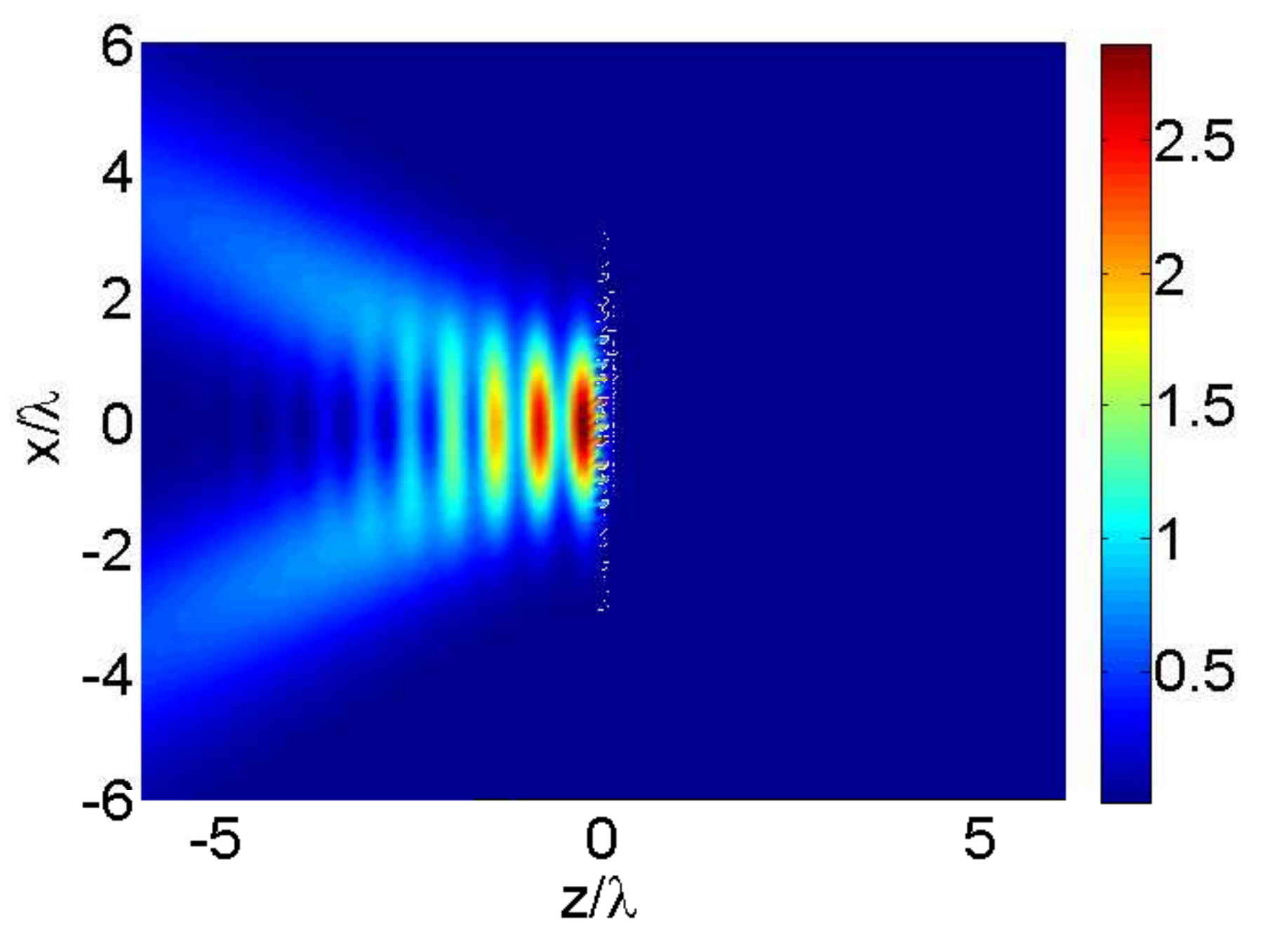}
\includegraphics[scale=0.4]{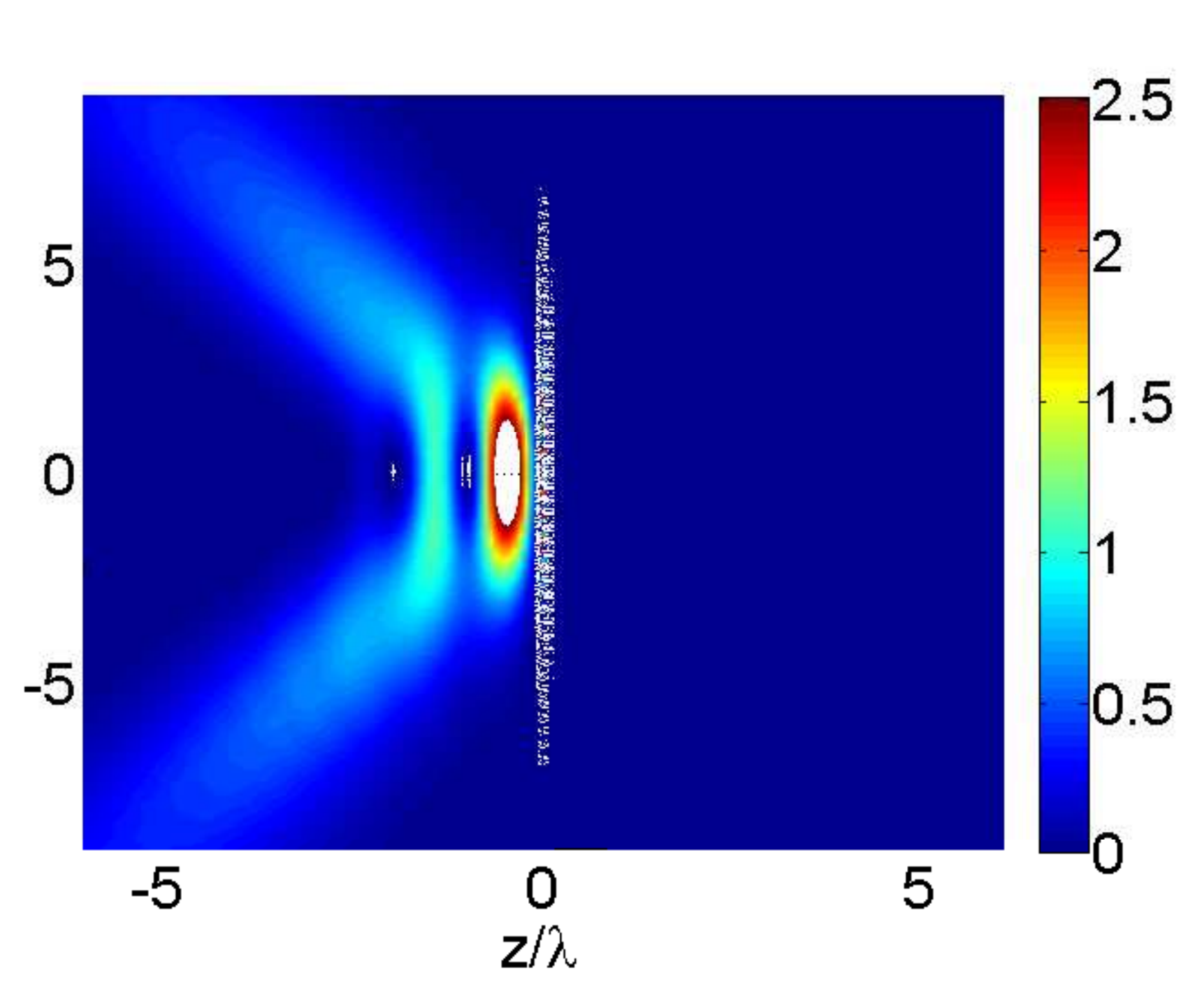}
\caption{\small{
Numerical approach, oblique incidence. (a) Reflection of a $p$-polarized Gaussian beam at an incident angle of $\theta=30^0$. The beam waist is $w_0=1.56\lambda$ and the array contains $40\times 40$ atoms with separation $a=0.2$. In accordance with the theoretical predictions (Fig. 3 in the main text), almost perfect reflection is observed. Quantitative agreement to the theory is found by examining the amplitude of the numerically calculated field along the $x$-axis at the far field, e.g. at $z=-6\lambda$ (noting the interference between the incident and reflected fields closer to the surface). (b) Same as (a) for $s$-polarized beam at an incident angle $\theta=60^o$. Due to the larger angle of incidence, a larger array of $70\times 70$ is taken, so that diffraction from the edges becomes very small and agreement to the theory (Fig. 3) is observed.
 }} \label{SM_fig4}
\end{center}
\end{figure}


\begin{thebibliography}{}
\bibitem{noteLS} The sum diverge due to the contribution of the Lamb shift whose counter term, the real part of $G_{ij}(0,0)$, was neglected here. In order to use these sums to obtain the real part of $g_{ij}$ we restore the real part of $G_{ij}(0,0)$, represent it by an integral in momentum space which has a similar divergence as the sum on $m$. Then, by regularizing both the sum and the integral with the same, e.g. exponential/Gaussian cutoff function, the difference of the two converges numerically and yields the same results obtained by the direct summation in real space, Eq. (\ref{SM_g}).
\end{thebibliography}
\end{document}